\documentclass[a4paper,11pt]{article}
\pdfoutput=1 

\usepackage{jcappub} 

\title{Baryon asymmetry and lower bound on right handed neutrino mass in fast expanding Universe: an analytical approach}

\author{Mainak Chakraborty,}
\author{Sourov Roy}

\affiliation{ School of Physical Sciences, Indian Association for the Cultivation of Science, 2A $\&$ 2B Raja S.C. Mullick Road, Kolkata 700 032, India }

\emailAdd{psmc2382@iacs.res.in}
\emailAdd{tpsr@iacs.res.in}

\abstract{ 
The expansion rate of the Universe deviates from its standard value when the total energy density includes contribution from a new scalar field apart from the radiation energy density. The non-trivial modifications incurred in the Boltzmann equations render the well known analytical solutions unsuitable in non standard scenario. In the present study we derive analytical expressions for the efficiency factor (which is nothing but solution of set of Boltzmann equations) using certain legible approximations. A fair degree of accuracy of these formulas have been observed by juxtaposing the analytical results with that obtained through numerical solution of Boltzmann equations. Faster expansion of the Universe results in decrement of the effective decay parameter which brings down the amount of washout of asymmetry due to inverse decay. Thus in non-standard cosmology scenario, a larger fraction of the asymmetry (generated at early epoch) is expected to survive till present epoch. Alteration of the cosmology does not affect the underlying particle physics model responsible for the generation of the CP asymmetry. Therefore starting from an identical particle physics model we will end up with a larger final baryon asymmetry in the non-standard scenario. It hints towards the possible relaxation of the lower bound of the lightest right handed neutrino mass required to produce adequate asymmetry which is in agreement with current experimental data. 
}

\makeatletter
\gdef\@fpheader{}
\makeatother

\begin{document}
\maketitle
\flushbottom

\section{Introduction}
Neutrino oscillation\cite{Pontecorvo:1967fh,Gribov:1968kq} experiments have now entered into a precision era. Several experiments are running at full throttle to minimize the ambiguities (mass ordering of light neutrinos, value of Dirac CP phase, nature of neutrinos:Dirac or Majorana, octant of the atmospheric mixing angle ) that still persist. Recent results of NO$\nu$A and T2K experiments seem to shed some light towards the possible range of values of CP violating phase $\delta$. A global analysis\cite{Esteban:2018azc} has been carried out incorporating the latest results of different neutrino
oscillation experiments like T2K\cite{Abe:2017vif,Abe:2018wpn}, NO$\nu$A\cite{Adamson:2017gxd,NOvA:2018gge}, Daya Bay\cite{Adey:2018zwh}, RENO\cite{Bak:2018ydk}, MINOS\cite{Adamson:2013whj}, Double CHooz\cite{Abe:2014bwa}. It has hinted towards Normal mass ordering (for Inverted ordering: $\Delta \chi^2=4.7$) and for the atmospheric mixing angle the second octant is preferred over the first octant ($\Delta \chi^2=4.5$). The best fit value of the Dirac CP phase turns out to be $215^\circ$ whereas the CP conserving case is disfavoured. The Majorana
nature of the neutrinos can only be confirmed if the neutrino less double beta decay is indeed observed. We only have upper bound on the concerned decay width as of now. Although  the experiments like EXO-200\cite{Anton:2019wmi}, GERDA\cite{Agostini:2018tnm}, KamLAND-Zen\cite{Gando:2018kyv}, CUORE\cite{Alduino:2017ehq}, CUPID\cite{Azzolini:2019tta} have not been able to detect any such decay yet, they are optimistic to have some decisive outcome in near the future.

On the other hand cosmological experiments like WMAP\cite{WMAP:2003elm,WMAP:2008lyn,WMAP:2008ydk,WMAP:2010qai,WMAP:2012nax} and Planck\cite{Planck:2018vyg,Planck:2013pxb,Aghanim:2016yuo} have ascertained the excess of matter over antimatter (in other word baryon over anti-baryon) in the Universe and its magnitude is also quoted within a certain error bar.  Theoretical explanation of this baryon asymmetry of the Universe(BAU) seems to be more natural and acceptable if it is assumed to be generated dynamically from a baryon symmetric universe (instead of treating this asymmetry as an initial condition). This dynamical process is commonly referred to as Baryogenesis\cite{Riotto:1998bt,Cline:2006ts,Dine:2003ax}.
Generation of baryon asymmetry through 
leptogenesis\cite{Fukugita:1986hr,Riotto:1999yt,Buchmuller:2004nz,Davidson:2008bu,Bertuzzo:2009im,DiBari:2012fz,Adhikary:2014qba,Samanta:2016hcj,Samanta:2018hqm,Samanta:2018efa} is most popular among the other alternatives(e.g Affleck-Dine\cite{Affleck:1984fy,Dine:1995kz} or GUT\cite{Ignatiev:1978uf,Ellis:1978xg} baryogenesis) and also it is easily relatable to the neutrino oscillation phenomenology. The process of leptogenesis naturally serves as a bridge between the cosmological matter-antimatter asymmetry and the physics of neutrino oscillation. Although the simplest Type-I seesaw mechanism\cite{Minkowski:1977sc,GellMann:1980vs,Yanagida:1980xy,Mohapatra:1979ia} (Standard Model+ 3 generation of heavy right handed (RH) neutrinos) is primarily used to explain the large leptonic mixing and tiny non-vanishing mass of active neutrinos, it can potentially produce the asymmetry in the lepton sector which is converted into baryon asymmetry through sphaleron transition.
The lepton number violating
(due to presence of Majorana mass term) out of equilibrium
decay of gauge singlet heavy RH neutrinos to lepton, Higgs pair creates a CP asymmetry which in turn produce the lepton asymmetry. The complex Yukawa coupling (between RH neutrino, lepton and Higgs) induces the CP violation
in the Lagrangian which is reflected in the unequal decay rate of the actual process and its CP conjugate one.
Thus the Sakharov\cite{Sakharov:1967dj} conditions of Baryogenesis are trivially satisfied here. 
The interrelation between the BAU and the Type-I seesaw mechanism has enabled us to constrain few seesaw parameters which were not yet bounded by oscillation data. One nice example of this is the so-called Davidson-Ibarra bound\cite{Davidson:2002qv}, which from the observed baryon asymmetry together with neutrino oscillation data sets a lower limit ($M_1\gtrsim 10^9$ GeV) on the mass of the decaying lightest RH neutrino (its mass spectrum has been chosen to be strongly hierarchical).
However there are few loop holes through which this bound can be evaded. One of them is the resonant 
leptogenesis\cite{Pilaftsis:1997jf, Pilaftsis:2003gt} scenario in which the resonantly enhanced CP asymmetry (adequate to meet the experimental bound ) is produced from the decay of nearly degenerate RH neutrinos of much lower mass (as low as a few TeVs). Flavoured leptogenesis\cite{Pilaftsis:2004xx,Pilaftsis:2005rv,Abada:2006fw,Abada:2006ea,Antusch:2006cw,Adhikary:2014qba,Samanta:2018hqm,Chakraborty:2020gqc} can also generate sufficient final baryon asymmetry (where the decaying RH neutrino is lighter than the prescribed lower limit) due to its decreased washout effects.
The baryon asymmetry is quantified in terms of a parameter $\eta_B$ which is the ratio of net baryon number density ($n_B - n_{\bar{B}}$) and photon density ($n_\gamma$). 
The baryon asymmetry parameter $\eta_B$ depends on efficiency factor $(\kappa)$ apart from the CP asymmetry parameter. The efficiency factor basically estimates what portion of asymmetry (which was generated at high temperature or early epoch through RH neutrino decay) survives down to low temperature or present epoch. The factor $\kappa$ is estimated by solving the corresponding Boltzmann equations for leptogenesis. Thus the final value (frozen value) of the baryon asymmetry parameter can be increased by enhancing the efficiency factor, while the CP asymmetry is kept constant.\\

In most of the existing literature the efficiency factor has been computed using standard radiation dominated cosmology, where the radiation energy density solely accounts for the total energy density of the Universe. Inclusion of any other component in the total energy content will change the expansion rate of the Universe, correspondingly the temperature dependence of the Hubble parameter gets modified which directly affects the solution of Boltzmann equations. It is to be mentioned that there have been several intriguing efforts to analyze the outcomes of deviation from the standard radiation dominated cosmology in pre-BBN era. Although there are several popular theories (such as quintessence model\cite{Fujii:1982ms,Ford:1987de,Chiba:1997ej,Salati:2002md,Profumo:2003hq,Tsujikawa:2013fta}, late decay of inflation field\cite{Arbey:2008kv,Arbey:2011gu}) in the existing literature to bring about the deviation from standard cosmology, in the present work we explore the simplest case where the non-standard effect in cosmology arises due to the presence of a new scalar field\cite{DEramo:2017gpl}.  \\

The field $\phi$ used in our work to drive the fast expansion of the Universe is a typical scalar field present in the cosmic fluid from post reheating era till an intermediate temperature (denoted by $T_R$ in our paper) which is obviously much higher than that of the BBN era. The total energy density of the Universe is dominated by the said scalar field in the temperature regime $T_R < T< T_{\rm reheat}$ and the standard radiation domination is gradually restored below $T_R$. The inflaton field decayed when the reheating phase sets in and therefore it(the inflaton) vanishes completely at the end of reheating. At this temperature regime the cosmic fluid should be abundant in the scalar $\phi$ if the bulk of total energy density is indeed contributed by $\phi$. Thus it can be said unambiguously that 
the scalar field $\phi$ itself is not identical with the inflaton. However the $\phi$ field can be produced from the decay\cite{DiMarco:2018bnw} of inflaton. We can find these type of scalar fields in 
scalar-tensor(ST) theory of gravity\cite{Catena:2009tm,Rehagen:2014vna,Meehan:2015cna,Catena:2004ba,Damour:1996xx} or string theory motivated orientifold models\cite{DiMarco:2018bnw,Angelantonj:2002ct}. One important aspect of these kind of theories is the absence of direct coupling of $\phi$ with SM fields. This scalar field $\phi$, has some non-standard coupling with gravity, can couple to the matter field only through the metric tensor. 
The nature of fast expansion,(or in other words to what extent the expansion rate deviates from the standard one) is characterized by the parameter $\delta$. A typical value of this expansion rate can be attributed to a specific structure of the scalar potential $V(\phi)$. Detailed discussion about the structure of these potential can be found in \cite{DEramo:2017gpl}.
\paragraph{}
Analytical solution of Boltzmann equations (within certain approximation) in the standard radiation dominated cosmology already exists in the literature\cite{Buchmuller:2004nz,Blanchet:2006dq,Samanta:2019yeg}. In our present work we have tried to find approximate analytical form of the efficiency factor $\kappa$ in a scenario where the cosmology deviates from its well known radiation dominated picture. Although there are a few recent works\cite{Chen:2019etb,Chang:2021ose,Konar:2020vuu,Mahanta:2019sfo} aiming to solve the Boltzmann equations numerically in modified cosmology, they have managed to show their results only for some benchmark values of the parameters. On the contrary, with analytical formulas of efficiency factor one can easily scan the whole range of parameters. Besides this one of our prime objectives is to check whether the efficiency factor is indeed enhanced in modified cosmology which may help us to evade the lower bound on RH neutrino mass (Davidson-Ibarra bound) without altering the value of CP asymmetry parameter\cite{Covi:1996wh}.  \\

To pave the way for a smooth transition from standard radiation dominated cosmology to the non-standard (or fast expanding) case the different sections of the manuscript are arranged systematically as follows. In 
Sec.\ref{lepto-standard-cosmo} we present a short recapitulation of standard radiation dominated cosmology. Its different subsections give a overview of the well known formulas of CP asymmetry in Type-I seesaw leptogenesis and analytical solution of the Boltzmann equations which were derived using standard cosmology. Sec.\ref{mod_cosmo} contains an exhaustive discussion about the modification of standard cosmology. Subsections under this section lead us from modification of expansion rate of Universe (Sec.\ref{mod-hubble}) to developing the Boltzmann equations (Sec.\ref{mod-boltz}) in the modified cosmology and finally analytical solutions (Sec.\ref{small_d_soln}, 
Sec.\ref{large_d_soln}) of these modified Boltzmann equations in non-standard cosmology. In Sec.\ref{deviation-bounds} we have analyzed whether non-standard cosmology has any impact on the existing bounds obtained from neutrino oscillation phenomenology and cosmological matter-antimatter asymmetry where special emphasis is given on the Davidson-Ibarra bound (Sec.\ref{mod-DI-bound}). The key features of the entire study have been summarized in 
Sec.\ref{summary}.     

\section{Leptogenesis in Standard cosmology}\label{lepto-standard-cosmo}
We first briefly review the formulas and related outcomes of hierarchical leptogenesis scenario in standard radiation dominated cosmology. Our main focus of this work is to find the solutions of Boltzmann equations, which provide the so called efficiency factor of leptogenesis. Since we are not studying any specific flavour symmetry and its implication on the CP asymmetry, the exact structure of the Yukawa matrices is not so important in this case. 
Nevertheless, in subsection (\ref{cp-asy}) we show the standard working formula for the CP asymmetry parameter and its highest possible value (supported by oscillation data) in terms of the mass of the decaying lightest RH neutrino. Evolution of this asymmetry is tracked by a set of Boltzmann equations which is directly connected to the expansion rate of Universe. In the subsection\ref{boltz-standard} we present a short discussion about the widely used Boltzmann equations for leptogenesis which are derived using standard cosmology. 

\subsection{CP asymmetry}\label{cp-asy}
The CP asymmetry produced due to the out of equilibrium decay of heavy RH neutrinos is basically the measure of imbalance in decay rate of \textquoteleft RH neutrino to lepton($l$), Higgs($H$) pair\textquoteright and its conjugate process. Considering the lepton flavours to be indistinguishable, the mathematical form of this CP asymmetry parameter due to decay of $i^{\rm th}$ RH neutrino is given by\cite{Covi:1996wh}
\begin{eqnarray}
\epsilon_i &=&\frac{\Gamma_{{N}_i\rightarrow
    l^-H^+,\nu_l H^0}-\Gamma_{{N}_i\rightarrow
    l^+H^-,\nu_l^c H^{0*}}}{\Gamma_{{N}_i\rightarrow
    l^-H^+,\nu_l H^0}+\Gamma_{{N}_i\rightarrow
    l^+H^-,\nu_l^c H^{0*}}}.
\label{cpasym}
\end{eqnarray}
The decay rates have been calculated taking into account the one loop self energy and vertex diagrams along with the tree level diagrams. 
Finally the CP asymmetry parameter can be expressed in terms of the model parameters as 
\begin{eqnarray}
\varepsilon_i=\frac{1}{8\pi v^2\left(m^\dagger_D m_D \right)_{ii}}
\sum_{j\ne  i}Im\left\{\left(m^\dagger_D m_D \right)_{ij}^2\right\}g(x_{ij})~,
\label{asy}
\end{eqnarray}
where $m_D$ is the Dirac neutrino mass matrix, related to the leptonic Yukawa $(Y)$ coupling matrices through SM Higgs VEV($v$) as $m_D=Yv$ and $g(x)$ is the loop function with its argument $(x)$ being the squared ratio of RH neutrino mass eigenvalues. The expression of $g(x)$ and its approximate form in the limit of hierarchical RH neutrinos $(x \gg 1)$  are as follows
\begin{eqnarray}
 g(x)&=&\sqrt{x}\left\{1-(1+x)\ln(1+\frac{1}{x})+\frac{1}{1-x}\right\}\\
     &\simeq& -\frac{3}{2\sqrt{x}}~~~~~~({\rm when}~x \gg 1)~.
\end{eqnarray}
The Type-I seesaw mechanism with three additional heavy RH neutrinos give rise to the effective light neutrino mass matrix,
\begin{equation}
m_\nu = -m_D {\hat{M}}^{-1}_R m^T_D ~,
\end{equation}
where $\hat{M}_R=diag(M_1,M_2,M_3)$ is the RH neutrino mass matrix, which is assumed to be in its mass basis. The effective light neutrino mass matrix $m_\nu$ is diagonalized through a unitary transformation $U$(popularly written in PMNS\cite{Maki:1962mu,Kobayashi:1973fv} parametrization) as
\begin{equation}
 U^\dagger m_\nu U^\ast =diag (m_1,m_2,m_3) =-\hat{m}_\nu~~~~~~({\rm say}), \label{diag_mnu}
\end{equation}
where $m_i(i=1,2,3)$ are three light neutrino mass eigenvalues. Using the Type-I seesaw relation in the L.H.S. of eq.\ref{diag_mnu} we get
\begin{eqnarray}
  U^\dagger m_D M^{-1}_R m^T_D U^\ast &=&\hat{m}_\nu ~.\label{diag_mnu1}
\end{eqnarray}
The diagonal $\hat{m}_\nu$ matrix can be absorbed in the L.H.S of the equation (\ref{diag_mnu1}) as a multiplicative factor of 
$\sqrt{\hat{m}^{-1}_\nu}$ (in extreme left and right of the expression). The resulting equation can be regarded as the orthogonality relation of a matrix $R$ given by 
\begin{equation}
R= \sqrt{M^{-1}_R} m^T_D U^\ast \sqrt{\hat{m}^{-1}_\nu} ~.\label{R_mat}
\end{equation}
The explicit functional dependence of the CP asymmetry parameters on the low energy parameters is easily understandable when the former is expressed in terms of the $R$ matrix\footnote{For this purpose we invert eq.\ref{R_mat} in order to express $m_D$ in terms of $R$ and $U$. This special type of parametrization is known as Casas-Ibarra\cite{Casas:2001sr} parametrization.}. A few steps of algebric manipulation lead us to 
\begin{eqnarray}
\epsilon_i &=&\frac{1}{8 \pi v^2 \sum\limits_{n^\prime} m_{n^\prime}| R_{i n^\prime} |^2} 
\sum_{j \neq i}  M_j  \sum_{n,k}m_n m_k~  Im \left \{ R_{jn} R_{jk} R_{in}^\ast R_{ik}^\ast \right\} g(x_{ij})~. \label{uflcp} \\
              & \simeq &- \frac{3 M_i}{16 \pi v^2 } \left[
                          \frac{\sum\limits_{k} m^2_k ~Im \left\{ {R^\ast}^2_{ik} \right\}}{\sum\limits_{n^\prime} m_{n^\prime}| R_{i n^\prime} |^2} \right] ~~~~({\rm for}~ x_{ij} \gg 1)~.
\end{eqnarray}
Using the orthogonality relation of $R$ ($\sum\limits_{k}R^2_{ki}=1$), it can be shown that the maximum value of the bracketed quantity of above equation is $(m_3-m_1)$. Thus the upper limit on the magnitude of the most relevant CP asymmetry parameter $\epsilon_1$\footnote{In strongly hierarchical RH neutrino mass spectrum, the asymmetry generated due the decay of heavier RH neutrinos $(N_2,N_3)$ is washed out during the inverse decay process. So the corresponding CP asymmetry parameters ($\epsilon_2$,$\epsilon_3$) are not important in this case.} is given by\cite{Davidson:2002qv}
\begin{eqnarray}
 |\epsilon_1| \leq \frac{3 M_1}{16 \pi v^2 } (m_3 -m_1)~.
\end{eqnarray}

\subsection{Boltzmann equations and their analytic solution}\label{boltz-standard}
Discussions of the previous sections have made it clear that the simple Type-I seesaw model has ample scope of CP  violation. Now the remaining task is to find how this asymmetry generated at some high temperature (of the order of mass of the decaying RH neutrino) evolves down to present day low temperature in a radiation dominated Universe. The microscopic evolution of phase space distribution ($f(p^\mu, x^\mu)$) of a particle species is governed by the Boltzmann equations, schematically which can be presented by in terms of Liouville operator (${\bf L}$\footnote{In its covariant form the 
Liouville operator\cite{Kolb:1990vq} is given by 
$\hat{{\bf L}} = p^\alpha \frac{\partial}{\partial x^\alpha} - \Gamma^\alpha_{\beta \gamma} p^\beta p^\gamma \frac{\partial}{\partial p^\alpha}$
where $\Gamma^\alpha_{\beta \gamma}$ is the Christoffel symbol.}) and collision operator (${\bf C}$) as\cite{Luty:1992un}
\begin{equation}
 \hat{{\bf L}} [f]=-\frac{1}{2}{\bf C}[f]. 
\label{boltz_eq1}
\end{equation}
Taking into account all possible interactions (decay and/scattering) which tend to change the number density of the particle under consideration and assuming standard FRW background the set of classical Boltzmann 
equations\cite{Pilaftsis:1997jf, Pilaftsis:2003gt,Adhikary:2014qba} are derived from above (\ref{boltz_eq1}) equation as
\begin{eqnarray}
\label{BEN_i} 
\frac{d \eta_{N_i}}{dz} &=& \frac{z}{H(z=1)}\ \bigg[\,\bigg( 1
\: -\: \frac{\eta_{N_i}}{\eta^{\rm eq}_{N_i}}\,\bigg)\, \bigg(\,
\Gamma^{D\; (i)} \: +\: \Gamma^{SY\; (i)} +\:
\Gamma^{SG\; (i)}\, \bigg) \bigg]\nonumber\\
                        &=& -\bigg \{ D_{i}(z) + D^{SY}_i(z) +D^{SG}_i(z) \bigg \} \bigg (\eta_{N_i}(z)-\eta^{\rm eq}_{N_i}(z)  \bigg), \\
  \label{BEL_i}
\frac{d \eta_L}{dz} &=& -\, \frac{z}{H(z=1)}\, \bigg[
\sum\limits_{i=1}^3\, \varepsilon_i\ 
\bigg( 1 \: -\: \frac{\eta_{N_i}}{\eta^{\rm eq}_{N_i}}\,\bigg)\, \bigg(\,
\Gamma^{D\; (i)} \: +\: \Gamma^{SY\; (i)} +\:
\Gamma^{SG\; (i)}\, \bigg) \nonumber\\ 
&&+\, \frac{1}{2}\, \eta_L\, \bigg\{ \sum\limits_{i=1}^3\, 
\bigg(\, \Gamma^{D\; (i)} \: +\: 
\Gamma^{WY\;(i)}\: 
+\: \Gamma^{WG\; (i)}\,\bigg)\:
\bigg\}\bigg]\, \nonumber \\
&=&-\sum\limits_{i=1}^3\, \varepsilon_i\ \bigg \{ D_{i}(z) + D^{SY}_i(z) +D^{SG}_i(z) \bigg \} \bigg (\eta_{N_i}(z)-\eta^{\rm eq}_{N_i}(z)  \bigg)\nonumber\\ && -\frac{1}{2}\eta_L \sum\limits_{i=1}^3
\bigg\{ \frac{1}{2} D_i(z) z^2 \mathcal{K}_2(z) +D^{WY}_{i}(z) + D^{WG}_{i}(z) \bigg\},
\end{eqnarray}
where $z=M_1/T$, $\eta$ denotes the number density of a particle species being scaled by the photon density. The subscripts of $\eta$, designates the particle species under consideration, i.e $N_i$ signifies $i$th RH neutrino whereas $L$ stands for lepton. $\mathcal{K}_2(z)$ stands for the modified Bessel function(2nd kind) of order 2. 
$\Gamma$ generically denotes decay and scattering cross sections (scaled by photon density) where the superscripts on it are used to differentiate between decay $(D)$ or scattering $(S)$. The other superscript indicate the nature of interaction, whether it is Gauge $(G)$ mediated or Yukawa $(Y)$ mediated. The various 
decay (or scattering) parameters  ($D_i$ parameters of eq.(\ref{BEN_i},\ref{BEL_i}) including decay and scattering) can be generically expressed as 
\begin{equation}
 D_i(z) =\frac{z}{H(z=1)}\frac{\Gamma^{D\;(i)}}{\eta^{\rm eq}_{N_i}(z)} ~.\label{Di}
\end{equation}
$H$ is the well known Hubble parameter which can also be seen as a measure of expansion rate of Universe. In standard radiation dominated universe $H$ is directly proportional to the square of the temperature $(T)$ (or equivalently inversely proportional to the square of redshift $z=M_1/T$). The actual analytical expression of $H$ taking into account all numerical factors is given by
\begin{equation}
H(T) \simeq \sqrt{\frac{8 \pi^3 g^\ast}{90}} \frac{T^2}{M_{\rm pl}} = 
\sqrt{1.66 g^\ast} \frac{M_1^2}{M_{\rm pl}}\frac{1}{z^2}. \label{hubble}
\end{equation}
Here we want to emphasize that the $z$ dependence of $H$ crucially depends on the components of total energy density of the Universe. The inverse square dependence (eq.(\ref{hubble})) may change if the total energy content of the Universe is comprised of components other than radiation. We will elaborately discuss on this issue and its possible implication on the set of Boltzmann equations in Sec.\ref{mod-boltz}. 

In the present work we will be considering leptogenesis from the decay of hierarchical RH neutrinos. Therefore it is sufficient to consider the contribution from lightest RH neutrino ($N_1$) which implies that Boltzmann equations have to be solved for one generation of RH neutrino only. Thus the generation index $i$ can be simply replaced by $1$ (or we may simply omit the index and use $N$ in place of $N_i$)  and the summation over it is no more necessary. Now neglecting the scattering terms (which will be subdominant in case of leptogenesis from the decay of hierarchical RH neutrinos) the Boltzmann equations in their simplest from can be written as\cite{Adhikary:2014qba,Chakraborty:2020gqc} 
\begin{eqnarray}
&& \frac{d \eta_{N}}{dz}=- D(z)  \bigg (\eta_{N}(z)-\eta^{\rm eq}_{N}(z)  \bigg) ~, \label{bn}\\
&&\frac{d \eta_{B-L}}{dz}= - \varepsilon_1 D(z) \bigg (\eta_{N}(z)-\eta^{\rm eq}_{N}(z)  \bigg) -
\eta_{B-L} \bigg\{ \frac{1}{4} D(z) z^2\mathcal{K}_2(z)  \bigg\} \nonumber\\
&&~~~~~~~~~=-\, \varepsilon_1 D(z) \bigg (\eta_{N}(z)-\eta^{\rm eq}_{N}(z)  \bigg) - W_{\rm ID}(z) \eta_{B-L} ~.\label{bl}
\end{eqnarray}
Using the definition of $D$ (eq.\ref{Di}) it can be represented in a convenient form as 
\begin{eqnarray}
D(z) = z \frac{\Gamma_{N}(T=0)}{H(z=1)} \frac{\mathcal{K}_1(z)}{\mathcal{K}_2(z)} 
     =z K\frac{\mathcal{K}_1(z)}{\mathcal{K}_2(z)},    \label{decay}
\end{eqnarray}
where $K= \frac{\Gamma_{N}(T=0)}{H(z=1)}$ is called the decay parameter corresponding to the lightest RH neutrino. In the second Boltzmann equation the term involving $W$ is the wash out term and since it is dominated
(for hierarchical RH neutrino spectrum)by the contribution of inverse decay, we are considering only the inverse decay in our simplistic approach. It is expressed In terms of the decay parameter $K$ as
\begin{equation}
W_{\rm ID}(z)= \frac{1}{4} z^2 \mathcal{K}_2( z) D(z) =\frac{1}{4} K z^3 \mathcal{K}_1( z) ~. \label{inv_decay}
\end{equation}
If we solve the set Boltzmann equations from early epoch ($T\rightarrow \infty ~{\rm or}~ z\rightarrow 0$) to present day low temperature ($T\rightarrow 0 ~{\rm or}~ z\rightarrow \infty$), we will notice that the $(B-L)$ (or $L$) asymmetry gets frozen around $z \sim 10$ and the asymmetry doesn't change significantly any more. Analytical 
solution\cite{Buchmuller:2004nz,Blanchet:2006dq,Samanta:2019yeg} of Boltzmann equations leads us to the final $(B-L)$ asymmetry as
\begin{equation}
\eta^f_{B-L}= \eta^{\rm in}_{B-L} ~e^{- \int \limits^{z\rightarrow \infty}_{z_{in}}
W_{\rm ID}(z^\prime) dz^\prime} -  \varepsilon_1 \kappa_f~ . 
\end{equation}
Absence of any pre-existing asymmetry ($\eta^{\rm in}_{B-L}=0$) gives a rather simpler formula, i.e
\begin{equation}
\eta^f_{B-L}=- \frac{3}{4}\varepsilon_1 \kappa_f ,
\end{equation}
where $\kappa_f$ is the final efficiency factor corresponding to $N_1$. Again this $B-L$ asymmetry parameter is related to the baryon asymmetry parameter $\eta_B$ by a numerical factor, i.e
\begin{equation}
\eta_B = \left(\frac{a_{sph}}{F} \right) \eta_{B-L}~,\label{conv}
\end{equation}
where, $a_{sph}(=28/79)$ is the sphaleron conversion factor and $F(=2387/86)$ is the dilution factor\cite{Buchmuller:2004nz}. The most recent experimental bound ($95\%$ CL) on $\eta_B$ has been provided by the Planck collaboration\cite{Planck:2018vyg} as
\begin{equation}
  6.29\times10^{-10}<\eta_B <6.46 \times10^{-10}~. \label{expt-asym}
\end{equation}

The efficiency factor is basically a systematic solution of Boltzmann equation and thus a function of $z$. The functional dependence on $z$ as well as the decay and scattering related parameters can be found by explicit evaluation of following double integral
\begin{equation}
 \kappa(z)= -\frac{4}{3}\int \limits^{z }_{z_{in}} d z^{\prime} \frac{d {\eta_N}(z^\prime)}{d z^{\prime}} ~
e^{- \int \limits^{z }_{z^\prime} W_{\rm ID}(z^{\prime\prime}) dz^{\prime\prime}} ~. \label{kz}
\end{equation}
The final value of the efficiency factor can be easily obtained from above equation by taking $z \rightarrow \infty$ (or some large enough numerical value in case of numerical solution). It is well known that in strong washout regime the RH neutrino number density achieves its equilibrium value very fast even if we start from vanishing initial abundance. So the approximation $\eta_N(z) \simeq \eta^{eq}_N(z)$ is valid approximately for the whole range of $z$. These approximations enables us to find the final efficiency factor as function of the decay parameter as\cite{Buchmuller:2004nz,Blanchet:2006dq,Samanta:2019yeg}
\begin{equation}
\kappa_f(K) \simeq \frac{2}{K z_B(K)} \bigg ( 1- e^{-\frac{K z_B(K)}{2}} \bigg ), \label{kpf}
\end{equation}
where $z_B$ is typical value of $z$, for which the 
integrand\footnote{$F(z^\prime) =\frac{d {\eta_N}(z^\prime)}{d z^{\prime}} ~e^{- \int \limits^{z }_{z^\prime} W_{\rm ID}(z^{\prime\prime}) dz^{\prime\prime}}$ is maximum at $z^\prime=z_B$} of eq.(\ref{kz}) shows a maxima. 
In strong washout regime an approximate analytical fit (which reproduces very close results to that of the actual numerical computations) for $z_B(K)$ has been found to be\cite{Blanchet:2006dq,Samanta:2019yeg}
\begin{equation}
z_B(K) \simeq 2+ 4 K^{0.13} ~ e^{-\frac{2.5}{K}} ~. \label{zB_standard}
\end{equation}

\section{Leptogenesis in modified cosmology} \label{mod_cosmo}
The amount of final (or equivalently frozen) asymmetry has a significant dependence on the expansion rate of the Universe, that is to say it is 
controlled by the Hubble parameter\footnote{It is to be noted that the Yukawa couplings have been assumed to be constant. Thus they do not have the control over the decay and washout parameters.} or more precisely the temperature ($T$) dependence of Hubble parameter. We expect a noticeable change in the final value of asymmetry when the Universe expands faster than the usual radiation dominated case. In the following sections we present a quantitative discussion on the modification of functional dependence of $H$ on $T$ (or $z$) and correspondingly its effect on the set of Boltzmann equations. Those modified equations are then solved analytically to bring out the functional form of the final efficiency factor. Accuracy of these newly introduced analytical formulas is then checked critically by comparing them with that of actual numerical solution of the modified Boltzmann equations.

\subsection{Modification of Hubble parameter}\label{mod-hubble}
In standard cosmology the pre-BBN era is assumed to be radiation dominated. Thus the total energy density of the Universe is composed of solely radiation energy, i.e
\begin{equation}
\rho \simeq \rho_{\rm rad} = \frac{\pi^2}{30} g_\ast (T) T^4 
\end{equation}
where $g_\ast (T)$ is total number of relativistic degrees of freedom at temperature $T$. Again using Friedmann's equation one can write 
\begin{equation}
H(T)=\sqrt{\frac{8 \pi G}{3}} {\rho (T)}^{1/2}. \label{hubble1}
\end{equation}
Now using the expression of Planck mass $(M_{\rm pl}= G ^{-1/2})$, the Hubble parameter in a radiation dominated cosmology can be simplified further as
\begin{equation}
H_{\rm rad}(T) =  \sqrt{\frac{8 \pi^3}{90} g_\ast(T)}\frac{T^2}{M_{\rm pl}} 
\simeq 1.66 \sqrt{g_\ast(T)} \frac{T^2}{M_{\rm pl}} ,
\end{equation}
or equivalently in terms of $z(=M_1/T)$,
\begin{equation}
H_{\rm rad}(z) = 1.66 \sqrt{g_\ast(T)} \frac{M_1^2}{M_{\rm pl}} \frac{1}{z^2} .
\end{equation}
We now examine how this $T$ (or $z$) dependence changes when some other component contributes to the total energy density along with the radiation. At early Universe we consider presence of another\cite{DEramo:2017gpl} scalar field $\phi$, whose energy density falls with the scale factor $(a(t))$ faster than that of radiation, i.e
\begin{equation}
\rho_\phi \sim a^{-(4+\delta)}, ~~{\rm with}~~ \delta > 0 ~~. 
\end{equation}
Let us consider $T_R$ to be a reference temperature at which the energy densities of the two energy components
become equal, i.e $\rho_{\rm rad}(T_R)=\rho_\phi(T_R)$. In the total energy density, for temperature greater than $T_R$\footnote{It is to be noted that $T_R$ is constrained by the BBN through the relation $T_R \gtrsim \left( 15.4\right)^{1/\delta}$ MeV}
\cite{Cyburt:2015mya,DEramo:2017gpl}, the contribution from $\phi$ field dominates, whereas below it the usual radiation energy density gradually takes over. 

The entropy density at a temperature $T$ is given by
\begin{equation}
s(T)= \frac{2\pi^2}{45}g_{\ast s}(T) T^3, 
\end{equation}
where $g_{\ast s}$ is the number of effective relativistic 
degrees of freedom\cite{Kolb:1990vq}. Now applying the conservation of co-moving entropy $(s(T)a(T)^3={\rm constant})$ we can connect the scale factors at temperatures $T$ and $T_R$ as
\begin{eqnarray}
 g_{\ast s}(T) T^3 \left(a(T)\right)^3 & = & g_{\ast s}(T_R) \left({T_R}\right)^3 \left( a(T_R)\right)^3, \nonumber\\
 \Rightarrow \frac{a(T_R)}{a(T)} & = & \frac{T}{T_R}\left[ \frac{g_{\ast s}(T)}{g_{\ast s}(T_R)} \right]^{1/3}.
\end{eqnarray}
This relation between the scale factors at two different temperatures enables us to calculate the ratio of energy densities of the $\phi$ field at the same pair of temperatures as
\begin{eqnarray}
\frac{\rho_\phi(T)}{\rho_\phi(T_R)} & = & \left[ \frac{a(T)}{a(T_R)} \right]^{-(4+\delta)} \nonumber\\
& = & \left[ \frac{g_{\ast s}(T)}{g_{\ast s}(T_R)}  \right]^{\frac{4+\delta}{3}} \left(\frac{T}{T_R} \right)^{4+\delta}. \label{phi_T_TR}
\end{eqnarray}
Now Considering the contribution of $\phi$ field along with the radiation the total energy density is given by
\begin{eqnarray}
\rho (T)= \rho_{\rm rad}(T) + \rho_\phi(T) = \rho_{\rm rad} (T) \left[ 1+ \frac{\rho_\phi(T)}{\rho_{\rm rad}(T)} \right]  ~.
\end{eqnarray}
Now replacing $\rho_\phi(T)$ by $\rho_\phi(T_R)$ using eq.(\ref{phi_T_TR}) the expression of total energy density can be rewritten as
\begin{eqnarray}
\rho(T)=\rho_{\rm rad}(T) \left[ 1+ \frac{\rho_\phi(T_R)}{\rho_{\rm rad}(T)}\left(  \frac{g_{\ast s}(T)}{g_{\ast s}(T_R)}\right)^{\frac{4+\delta}{3}} \left(\frac{T}{T_R} \right)^{4+\delta}  \right] ~.
\end{eqnarray}
Using the equality of the two energy components ($\phi$ and radiation) at $T=T_R$ the ratio $\rho_\phi(T_R)/\rho_{\rm rad}(T)$ can be expressed in terms of corresponding temperatures and as a result the expression of total energy density can be presented in a more convenient form as
\begin{equation}
\rho (T)  =\rho_{\rm rad}(T) \left[ 1+ \frac{g_\ast(T_R)}{g_\ast(T)} \left(  \frac{g_{\ast s}(T)}{g_{\ast s}(T_R)}\right)^{\frac{4+\delta}{3}} \left(\frac{T}{T_R} \right)^{\delta}  \right] ~.
\end{equation}
Using eq.(\ref{hubble1}) the Hubble parameter in this case can be expressed as
\begin{eqnarray}
 H(T)   & = & \sqrt{\frac{8 \pi G \rho_{\rm rad}(T)}{3}}   \left[ 1+ \frac{g_\ast(T_R)}{g_\ast(T)} \left(  \frac{g_{\ast s}(T)}{g_{\ast s}(T_R)}\right)^{\frac{4+\delta}{3}} \left(\frac{T}{T_R} \right)^{\delta}  \right]^{1/2} \nonumber \\
        & = & H_{\rm rad} (T) \left[ 1+ \frac{g_\ast(T_R)}{g_\ast(T)} \left(  \frac{g_{\ast s}(T)}{g_{\ast s}(T_R)}\right)^{\frac{4+\delta}{3}} \left(\frac{T}{T_R} \right)^{\delta}  \right]^{1/2}
\end{eqnarray}
It is to be noted that in the present work the masses of the decaying RH neutrinos can vary from few TeVs to the scale of grand unification. Therefore the temperature range we are concerned about is more or less same as the mass range mentioned above. Throughout the said temperature range the effective relativistic degrees of freedom\footnote{Both $g_\ast(T)$ and $g_{\ast s}(T)$} remains nearly constant. This assumption leads us to a very simplified form of Hubble parameter i.e
\begin{equation}
H(T)=H_{\rm rad}(T) \left[ 1+ \left(\frac{T}{T_R} \right)^{\delta} \right]^{1/2},
\end{equation}
which can also be expressed in terms of $z(=M_1/T)$ as
\begin{equation}
H(z) =H_{\rm rad}(z) \left[ 1+ \left(\frac{\gamma}{z} \right)^{\delta} \right]^{1/2}, \label{hubble_mod}
\end{equation}
where we have introduced a new parameter $\gamma = M_1/T_R$ with $M_1$ being the mass of the lightest RH neutrino. The presence of the $\phi$ field has modified the expansion rate of Universe noticeably which is manifested in the new expression of Hubble parameter (eq.\ref{hubble_mod}) which clearly differs from its radiation dominated counter part through a $z$ dependent function. The new Hubble $H(T)$ merges with 
$H_{\rm rad}(T)$ for very large values of $z$ (or equivalently at low temperatures). We will now check what changes are inflicted in the set of Boltzmann equations through this  modified Hubble parameter(eq.\ref{hubble_mod}).

\subsection{Boltzmann equations in the modified cosmological scenario}\label{mod-boltz}
Let us now recall the simplest form of Boltzmann equations (eqs.\ref{bn},\ref{bl}) in radiation dominated cosmology where the decay and inverse decay parameters are defined through eqs.(\ref{decay},\ref{inv_decay}). In faster expanding Universe the structure of the equations remains unaltered, whereas the decay and inverse decay parameters gets modified as
\begin{eqnarray}
D^{\prime}(z) & = & \frac{\Gamma(z)}{H(z)z} \nonumber \\
     & = & z \frac{\Gamma(z\rightarrow \infty)}{H_{\rm rad}(z=1)} \left[ 1+ \left(\frac{\gamma}{z} \right)^{\delta} \right]^{-1/2} \frac{\mathcal{K}_1(z)}{\mathcal{K}_2(z)} \nonumber\\
     & = & z K \left[ 1+ \left(\frac{\gamma}{z} \right)^{\delta} \right]^{-1/2}\frac{\mathcal{K}_1(z)}{\mathcal{K}_2(z)}  \nonumber \\
     & = & z K_{\rm eff}(z) \frac{\mathcal{K}_1(z)}{\mathcal{K}_2(z)}, \label{mod_D}
\end{eqnarray}
where the effective decay parameter is given by
\begin{equation}
 K_{\rm eff}(z) =K \left[ 1+ \left(\frac{\gamma}{z} \right)^{\delta} \right]^{-1/2}
 =K f(z), \label{Keff}
\end{equation}
  where $f(z)=\left[ 1+ \left(\frac{\gamma}{z} \right)^{\delta} \right]^{-1/2}$.The Decay parameter of the modified cosmology gets back its value of radiation dominated era for $z \rightarrow\infty$. The modified decay term (eq.\ref{mod_D}) seems to be identical to its radiation dominated counterpart, apart from the difference that the decay parameter ($K$) is now replaced by a $z$ dependent parameter $K_{\rm eff}(z)$. Similar argument follows for inverse decay term, which is now expressed as
\begin{equation}
 W^{\prime}_{\rm ID}(z)= \frac{1}{4} K_{\rm eff}(z) z^3 \mathcal{K}_1( z)
 = W_{\rm ID}(z) f(z)~. \label{mod_inv_decay}
\end{equation}
After finding the exact functional from of the decay and inverse decay terms we are now in a position to proceed for the analytical solution of Boltzmann equations\footnote{i.e we have to solve eq.\ref{bn}, eq.\ref{bl} with decay and inverse decay terms of eq.\ref{mod_D} and eq.\ref{mod_inv_decay} respectively} in
modified cosmological scenario.

\subsection{An approach towards analytical solution of new set of Boltzmann equations} \label{small_d_soln}
Before entering into the mathematical rigor of the solution of Boltzmann equations, let us recall the definitions\cite{Buchmuller:2004nz} of a few key parameters related to this solution. In case of thermal leptogenesis, even if we start with the assumption of vanishing initial abundance of RH neutrinos, they are created in the thermal bath due to the inverse decay of lepton Higgs pair. Eventually the RH neutrino abundance $(\eta_{N}(z))$ reaches its equilibrium value $(\eta^{\rm eq}_{N}(z))$ at a specific value of $z$ denoted as $z_{eq}$. In strong washout regime $(K >>1)$ this equilibrium is attained very fast(i.e $z_{eq} <<1$), whereas for weak washout case
it takes considerable time(i.e $z_{eq} >>1$). For $z>z_{eq}$ the $N_1$ abundance curve nearly follows the
equilibrium curve in strong washout regime. On the contrary weak washout regime exhibits a bit difference where downfall of $N_1$ abundance sets in at some later time (compared to that of equilibrium curve). \\

It can be easily understood from the double integral of eq.\ref{kz} that the efficiency factor $\kappa(z)$ is made up of a positive $(\kappa^+)$ and a negative $(\kappa^-)$ contribution. The positive slope of $N_1$ abundance curve $({\rm for}~z<z_{eq})$ gives rise to the negative part while contribution from the region beyond $z_{eq}$ turns out to be positive due to negative slope of the concerned curve. It can be expressed in a concise manner as\cite{Buchmuller:2004nz}
\begin{equation}
 \kappa(z) = \kappa^-(z<z_{eq}) + \kappa^+(z>z_{eq})~.
\end{equation}
In case of strong washout, since $z_{eq}<<1$ the corresponding negative contribution $\kappa^-$ turns out to be negligible compared to the positive contribution which allows us to take 
$ \kappa \simeq \kappa^+$. \\

\subsubsection{Analytical expression for the positive contribution $\kappa^+(z>z_{eq})$}
We first concentrate on the calculation of the efficiency factor in the strong washout regime. The double integral of eq.\ref{kz} is now modified as
\begin{equation}
\kappa(z) \simeq \kappa^+(z) = -\frac{4}{3}\int \limits^{z }_{z_{in}>z_{eq}} d z^{\prime} \frac{d {\eta_N}(z^\prime)}{d z^{\prime}} ~ e^{- \int \limits^{z }_{z^\prime} W^{\prime}_{\rm ID}(z^{\prime\prime}) dz^{\prime\prime}}  ~. \label{kz_mod}
\end{equation}
The slope of the $\eta_N(z^{\prime})$ curve can be approximated\footnote{It is based on the implicit assumption that $\eta_N$ follows equilibrium curve from very small value of $z$ (since $z_{eq}<<1$).} as 
$\frac{d {\eta_N}(z^\prime)}{d z^{\prime}} \simeq \frac{d {\eta^{eq}_N}(z^\prime)}{d z^{\prime}} =
-\frac{3}{2 K_{eff}(z^\prime) z^\prime } W^{\prime}_{\rm ID}(z^{\prime})$ which on substitution in the above
equation gives
\begin{eqnarray}
\kappa^+(z) & = & \frac{2}{K} \int \limits^{z }_{z_{in}} d z^{\prime} \frac{W^{\prime}_{\rm ID}(z^{\prime})}{f(z^\prime)z^\prime} ~ e^{- \int \limits^{z }_{z^\prime} W^{\prime}_{\rm ID}(z^{\prime\prime}) dz^{\prime\prime}} \nonumber \\
          & = & \frac{2}{K \bar{z}} \int \limits^{z }_{z_{in}} d z^{\prime}\frac{1}{f(z^\prime)}
          \frac{\bar{z} W^{\prime}_{\rm ID}(z^{\prime})}{z^\prime} ~ 
          e^{- \int \limits^{z }_{z^\prime} W^{\prime}_{\rm ID}(z^{\prime\prime}) dz^{\prime\prime}} 
          \nonumber\\
          &=& \frac{2}{K \bar{z}} \int \limits^{z }_{z_{in}} d z^{\prime}\frac{1}{f(z^\prime)}
           \overline{W}^{\prime}_{\rm ID}(z^{\prime}) ~ 
          e^{- \int \limits^{z }_{z^\prime} W^{\prime}_{\rm ID}(z^{\prime\prime}) dz^{\prime\prime}} 
\end{eqnarray}
where we have redefined\cite{Buchmuller:2004nz} the inverse decay term as $\overline{W}^{\prime}_{\rm ID}(z^{\prime})=\frac{\bar{z}}{z^\prime}W^{\prime}_{\rm ID}(z^{\prime})$. Here $\bar{z}$ is a specific value of $z$ represented mathematically as $\bar{z}={\rm Min}\{ z_f, z_B \}$\footnote{It is clear that if we want to estimate the efficiency factor at $z>z_B$, $\bar{z}=z_B$ can be used without any ambiguity.} where $z_f$ is the upper limit of the integration and at $z_B$ the integrand shows a distinct maxima. In this context we would like to clarify that numerical value of this $z_B$ is different from that of the standard cosmology and obviously its analytical expression can no longer be approximated as eq.\ref{zB_standard}.  \\

We have also checked that the integrand shows a sharp maxima around $z_B$ and the curve falls rapidly even for a small deviation from $z_B$. The integration basically represents the area under the curve (from $z_{in}$ to $z$). The special functional form of the integrand ensures that the maximum contribution to the whole integration comes from a very narrow region around $z_B$. Thus there won't be any noticeable change in the result of the integration if the $W^{\prime}_{\rm ID}$ of the exponent is replaced by $\overline{W}^{\prime}_{\rm ID}$. With this replacement the integral can be written in the form of a total derivative as
\begin{equation}
 \kappa^+(z) = \frac{2}{K \bar{z}} \int \limits^{z }_{z_{in}} d z^{\prime}\frac{1}{f(z^\prime)}
 \frac{d}{d z^\prime}\left( e^{- \int \limits^{z }_{z^\prime} \overline{W}^{\prime}_{\rm ID}(z^{\prime\prime}) dz^{\prime\prime}} \right).
\end{equation}
We have examined that around the point $z_B$ the variation of $f(z)$ is very marginal 
i.e $|f(z_B \pm \Delta z)-f(z_B)|<<f(z_B)$ ($\Delta z$ is small variation in $z$ around $z_B$). This allows us to treat $f(z)$ as constant with a value $f(z_B)$ through out the important range of $z$ ($z_B \pm \Delta z$) where the integral receives maximum contribution. We are now free to take $f(z_B)$ out of integration which makes the integration even simpler, i.e
\begin{eqnarray}
 \kappa^+(z) & = & \frac{2}{K \bar{z}}\frac{1}{f(z_B)} \int \limits^{z }_{z_{in}} d z^{\prime}
 \frac{d}{d z^\prime}\left( e^{- \int \limits^{z }_{z^\prime} \overline{W}^{\prime}_{\rm ID}(z^{\prime\prime}) dz^{\prime\prime}} \right)~, \nonumber \\
           & = & \frac{2}{K \bar{z}}\frac{1}{f(z_B)}  \left[ 1- e^{- \int \limits^{z }_{z_{in}} \overline{W}^{\prime}_{\rm ID}(z^{\prime\prime}) dz^{\prime\prime}} \right]~.
 \end{eqnarray}
Using the approximation of constant $f(z)$, the integration\footnote{Detailed calculation is given in appendix\ref{wid_int}} of the exponent is carried out from very small value of $z$ to a arbitrary value (which is assumed to be greater than $z_B$). It yields a very simple result
\begin{equation}
\int \limits^{z }_{z_{in} \rightarrow 0} \overline{W}^{\prime}_{\rm ID}(z^{\prime\prime}) dz^{\prime\prime} 
= \frac{f(z_B) K z_B}{4}\left( 2 -z^2 \mathcal{K}_2(z) \right)~,
\end{equation}
using which the simplified analytic form of the efficiency factor becomes
\begin{equation}
 \kappa^+(z) = \frac{2}{K z_B}\frac{1}{f(z_B)}  
 \left[ 1- e^{-\frac{f(z_B) K z_B}{4}\left( 2 -z^2 \mathcal{K}_2(z) \right) } \right] ~. \label{kz_mod1}
\end{equation}
\subsubsection{Analytical expression for the negative contribution $\kappa^-(z<z_{eq})$} \label{kp_neg_small_d}
Assuming dynamical initial abundance it has already been shown in ref\cite{Buchmuller:2004nz} that the negative part of the efficiency factor is directly connected to the instantaneous RH neutrino abundance $(\eta_N(z))$ through the relation
\begin{equation}
 \kappa^-(z) = -2 \left( 1 - e^{-\frac{2}{3}\eta_N(z)} \right)~.\label{kp_neg}
\end{equation}
Although this formula is appropriate for the present case (i.e modified cosmology), the analytical expression for $\eta_N(z)$ differs from that of the radiation dominated case\footnote{It has been shown in Ref:\cite{Buchmuller:2004nz} that $N_1(z)$ abundance in strong washout regime is given by $\eta_N(z)=\frac{3}{4} \left( 1- e^{-\frac{K z^3}{6}} \right)$ (when $z<z_{eq}$).}. Imprint of modified cosmology can be observed in the new expression for RH neutrino abundance through the parameters $\gamma$ ans $\delta$ as (derivation presented in appendix\ref{eta_N_mod})
\begin{equation}
 \eta_N(z)=\frac{3}{4} \left[ 1- \exp \left({-\frac{K \gamma^{\frac{\delta}{2}} z^{\frac{6+\delta}{2}}}{6+\delta}}\right) \right]~. \label{mod_N}
\end{equation}
This analytical formula for $\eta_N(z)$ successfully reproduces the correct numerical value which perfectly matches with $\eta_N(z)$ evaluated through direct numerical solution of corresponding Boltzmann 
equation (eq.\ref{BEN_i}) with modified Decay term (eq.\ref{mod_D}). It is to be noted that the above mentioned analytical formula (eq.\ref{kp_neg}) for $\kappa^-$ works perfectly well before $\eta_N(z)$ reaches equilibrium. Since we are working in strong washout regime, the asymmetry produced at early epoch ($z<<1$ or $T>M_1$) will be erased significantly at high $z$. Evolution of this asymmetry is represented by 
$\kappa^-(z)$. For strong washout this contribution can be safely neglected for $z>z_{eq}$. 

\subsubsection{Complete expression of the efficiency factor $\kappa(z)$} \label{kp_total_small_d}
The total efficiency factor at any point $z$ should be computed by algebraically adding the positive($\kappa^+$)  and negative($\kappa^-$) contribution. However, it has to be kept in mind that significant contribution from $\kappa^-$ arises for $z<z_{eq}$ whereas $\kappa^+$ is valid only for $z>z_{eq}$ by its definition. Therefore total efficiency factor ($\kappa$) is effectively described solely by $\kappa^-$ for 
$z<z_{eq}$ and by $\kappa^+$ for $z>z_{eq}$. The total efficiency factor can be expressed in a compact mathematical notation as 
\begin{eqnarray}
\kappa (z) & \simeq & \kappa^+(z) \Theta (z-z_{eq})  + \kappa^-(z) \Theta (z_{eq} -z) \nonumber \\
           &   = & \frac{2}{K z_B f(z_B)}
 \left[ 1- e^{-\frac{f(z_B) K z_B}{4}\left( 2 -z^2 \mathcal{K}_2(z) \right) } \right] \Theta (z-z_{eq})  
 -2 \left( 1 - e^{-\frac{2}{3}\eta_N(z)} \right) \Theta (z_{eq} -z)~, \nonumber \\  \label{kp_tot}
\end{eqnarray}
where $\Theta$ is a step function defined as 
\begin{eqnarray}
&&\Theta (x) =0 ~~~~{\rm for ~x<0} \nonumber\\
&&\Theta (x) =1 ~~~~{\rm for ~x>0} ~.\nonumber\\
\end{eqnarray}
The final value of efficiency factor can be computed from eq.\ref{kp_tot}  by taking a large enough value of $z$, where the negative contribution ($\kappa^-$) is already absent and the positive term ($\kappa^+$) is further simplified as the second term of the exponent vanishes, i.e $\left[ z^2 \mathcal{K}_2(z)\right]_{z\rightarrow \infty} = 0$. It leads to an expression of final efficiency factor as
\begin{equation}
 \kappa_f = \kappa(z \rightarrow \infty) \simeq \kappa^+(z \rightarrow \infty) = \frac{2}{K z_B}\frac{1}{f(z_B)}  
 \left[ 1- e^{-\frac{f(z_B) K z_B}{2} } \right] ~. \label{kf_mod1}
\end{equation}
It is worthwhile to mention that the above derivation implicitly relies on the assumption that our range of integral $(z_{in} ~{\rm to}~z)$ always includes the point $z_B$. This treatment will utterly fail if $z_B$ falls out side the integration limit. Since we are working in the strong washout regime where 
$z_{eq}$ is always very small, the condition $z_{in}<z_B$ is automatically satisfied. 
%
\subsubsection{Reliability of the analytical formulas} \label{plots_small_d}
\begin{figure}[h!]
\begin{center}
\includegraphics[width=5.5cm,height=5cm,angle=0]{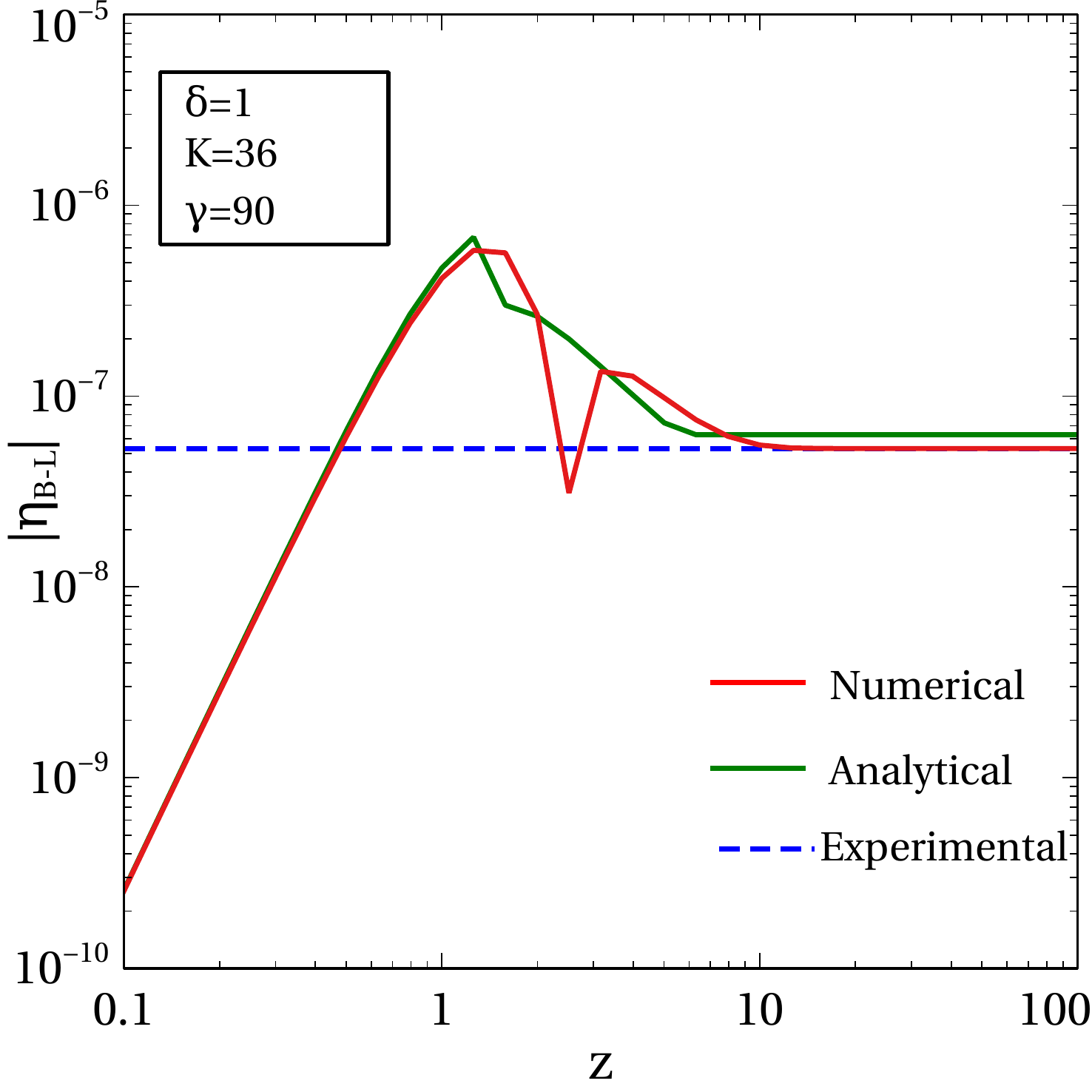}
\hspace{.5cm}
\includegraphics[width=5.5cm,height=5cm,angle=0]{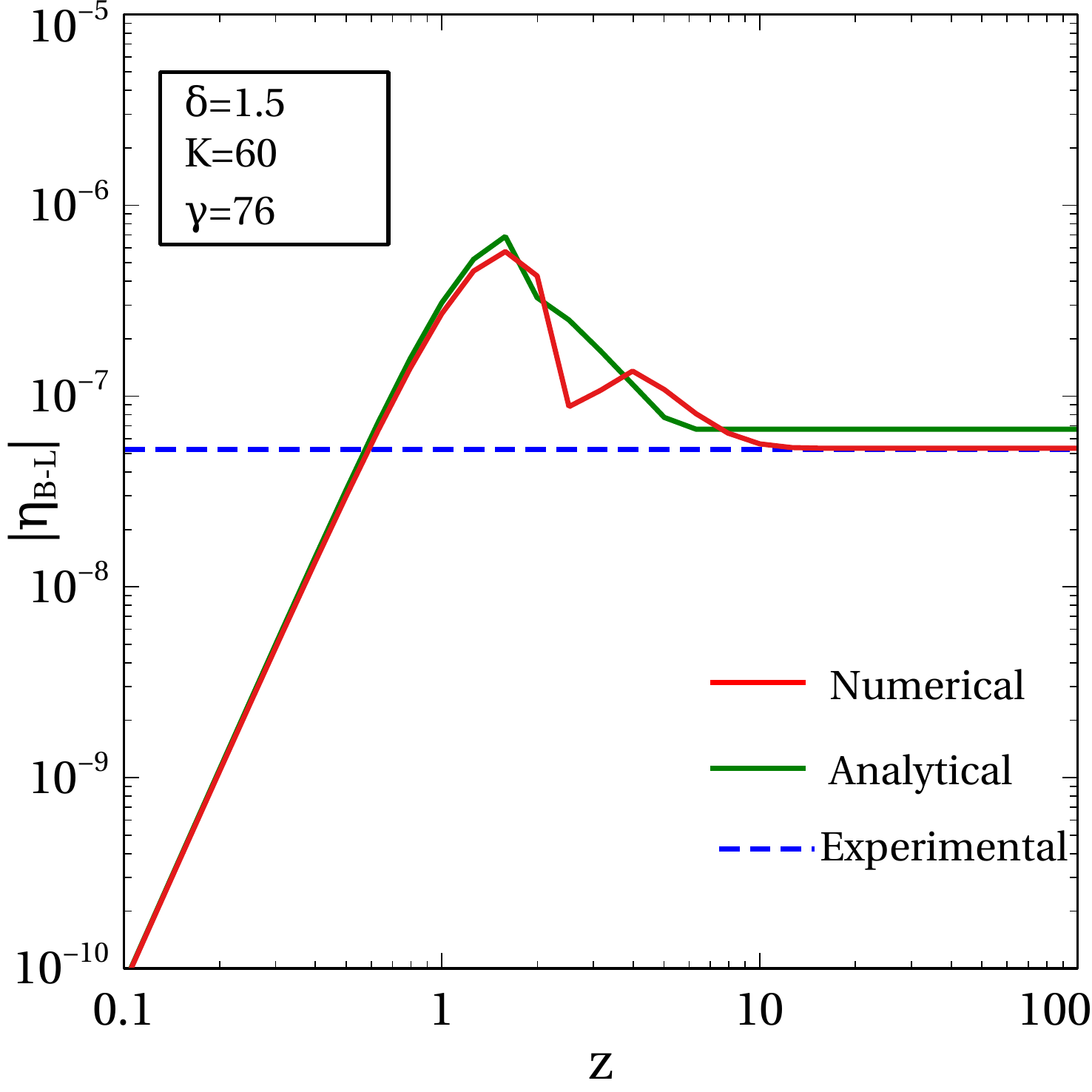}
\caption{$\eta_{B-L}(z)$ evaluated through numerical and analytical solution of Boltzmann equations for $\delta=1$(left) and $\delta=1.5$(right). A fixed value of the CP asymmetry parameter $\epsilon_1=-2\times10^{-6}$ has been used for all the cases. 
Using the experimental bound(eq.\ref{expt-asym}) on $\eta_B$ in eq.\ref{conv}, the final value of $B-L$ asymmetry can be constrained as $5.24\times10^{-8}<|\eta_{B-L}(z\rightarrow\infty)|<5.38\times10^{-8}$ ($95\%$ CL). The allowed range of 
$\eta_{B-L}$ is represented by the horizontal blue dashed line.
Analytically calculated final asymmetry ($\eta_{B-L}(z\rightarrow\infty)$) shows fair agreement with that of the numerical solution.}
\label{del1_2}
\end{center}
\end{figure}

\begin{figure}[h!]
\begin{center}
 \includegraphics[width=5.5cm,height=5cm,angle=0]{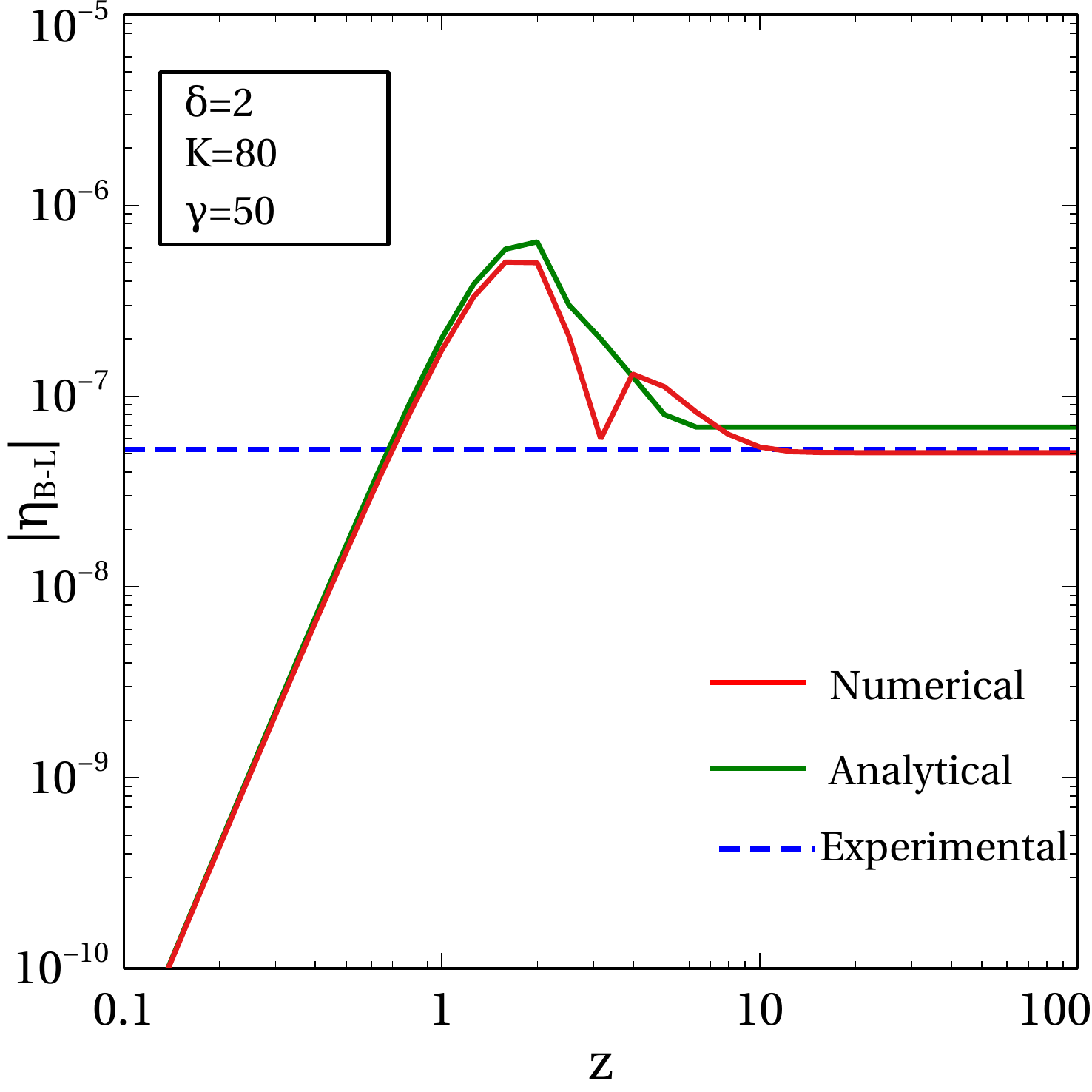}
\hspace{.5cm}
\includegraphics[width=5.5cm,height=5cm,angle=0]{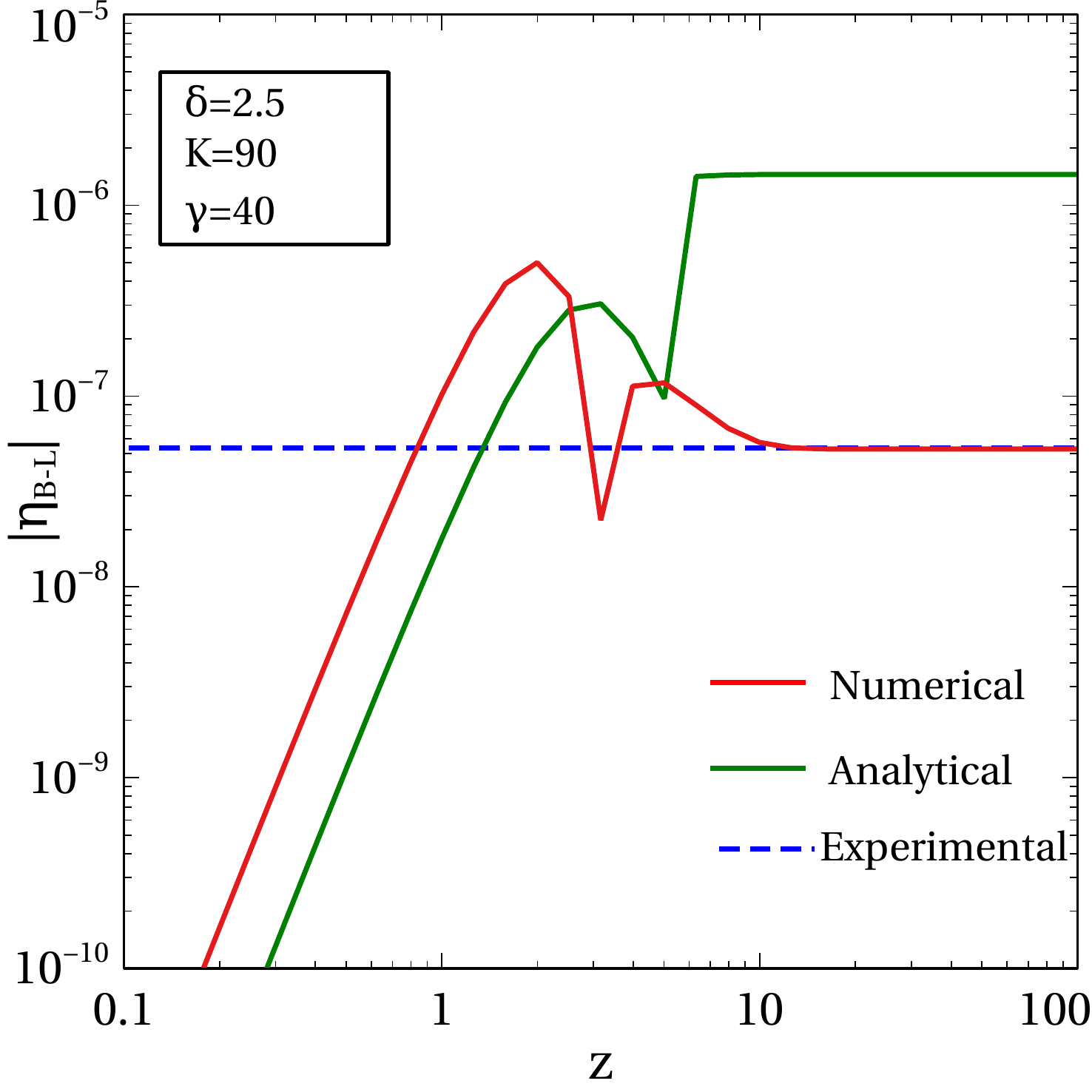}
\caption{$\eta_{B-L}(z)$ evaluated through numerical and analytical solution of Boltzmann equations for $\delta=2$(left) and $\delta=2.5$(right). For $\delta=2.5$ the numerical and the analytical solutions differ significantly from each other. }
\label{del2.5_3}
\end{center}
\end{figure}

We now examine the validity of the analytical expression of efficiency factor $\kappa(z)$ (eq.\ref{kp_tot}) for different values of $\delta$ through graphical representations. 
The $B-L$ asymmetry parameter (i.e, $\eta_{B-L}(z)$) has been calculated once solely by numerical computation and then using analytical formula. The corresponding results are represented graphically in the same plot. This process is repeated four times for four different values of 
$\delta$ (left and right panel of Fig.\ref{del1_2} and Fig.\ref{del2.5_3} ).


Ideally the analytically generated curve should merge with its numerically produced counterpart. We notice that the flat portion ($\eta_{B-L}(z \rightarrow \infty)$) of the curves shows excellent agreement only up to $\delta \sim 2$. The analytical result start to differ from numerical result as soon as $\delta \gtrsim 2$ and this difference becomes more and more significant with increase of 
$\delta$ hinting towards incapability of the analytical formulas in reproducing correct asymmetry for faster expansion 
(i.e higher\footnote{To be specific `lower' (`higher') implies $\delta \lesssim 2$ 
($\delta \gtrsim 2$). } values of $\delta$).  \\

Using the experimental bound on $\eta_B$ in eq.\ref{conv}, the allowed range of $\eta^f_{B-L}$ comes out to be 
($5.24\times10^{-8}<\eta^f_{B-L}<5.38\times10^{-8}$), which is represented by a thick horizontal blue dashed line in the plots. It should be emphasized that we have chosen some specific values the set of free parameters (indicated inside the box present in the plots) such that the final value of baryon asymmetry agrees with its experimental counterpart.  
\subsection{In search of refined analytical formulas valid for faster expansion}\label{large_d_soln}
The discussions and supporting graphical representations of the previous section has firmly established the validity of analytic expressions of efficiency factors for lower values of $\delta$. Nevertheless, this excellent agreement with the numerical solution of Boltzmann equation ceases to hold as soon as $\delta$ exceeds $2$. 
This situation suggests towards the invalidity of earlier approximations (which were used to derive eq.\ref{kz_mod1} and eq.\ref{kf_mod1}) in the present scenario. We now analyze where exactly the approximations fails and subsequently refine the analytical formulas without any approximations or at least use such approximations which are fit for current situation (i.e $\delta \gtrsim 2$). \\

This sharp contrast between regions of lower and relatively higher values of $\delta$ needs to be clarified. A deeper understanding about the interrelation between $\delta$ and $z_B,z_{eq}$ may guide us to unearth the underlying reason for the failure of previous approximations/assumptions. An approximate analytical expression of $z_{eq}$ has been found to be\footnote{The derivation is given in appendix\ref{zeq_mod}}
\begin{equation}
z_{eq} =\left( \frac{p \gamma^{\delta/2}}{K}  \right)^\frac{2}{2+\delta} , \label{zeq}
\end{equation}
where $p$ is a numerical factor and in our case $p \sim 10$ shows fair agreement with that of the numerical solution. It would be easier to perceive the connection between `validity of the approximations' and $z$ dependence of  $z_{eq}, z_B$ through proper pictorial representations. In the left panel of the following figure (fig.\ref{zb_zeq}) we have shown the evolution of $N_1$ abundance obtained by numerical solution of first Boltzmann equation for different values of $\delta$ while the other parameters $\gamma$ and $K$ have been kept fixed. The equilibrium abundance ${\eta^{eq}_N}(z)$ has been shown in the same plot. The $\eta_N(z)$ curves (for different values of $\delta$) touches its equilibrium counterpart at different values of abscissa, which actually denotes the values for $z_{eq}$ for corresponding values of $\delta$. In the right panel we plot $z_B(\delta)$ and $z_{eq}(\delta)$\footnote{This $z_{eq}$ has been calculated using the analytical formula of eq.\ref{zeq} } simultaneously. So the viability of the analytical formula for $z_{eq}$ may be ascertained through comparison between the left and right panel of fig.\ref{zb_zeq}.
\begin{figure}
\begin{center}
\includegraphics[width=6cm,height=5cm,angle=0]{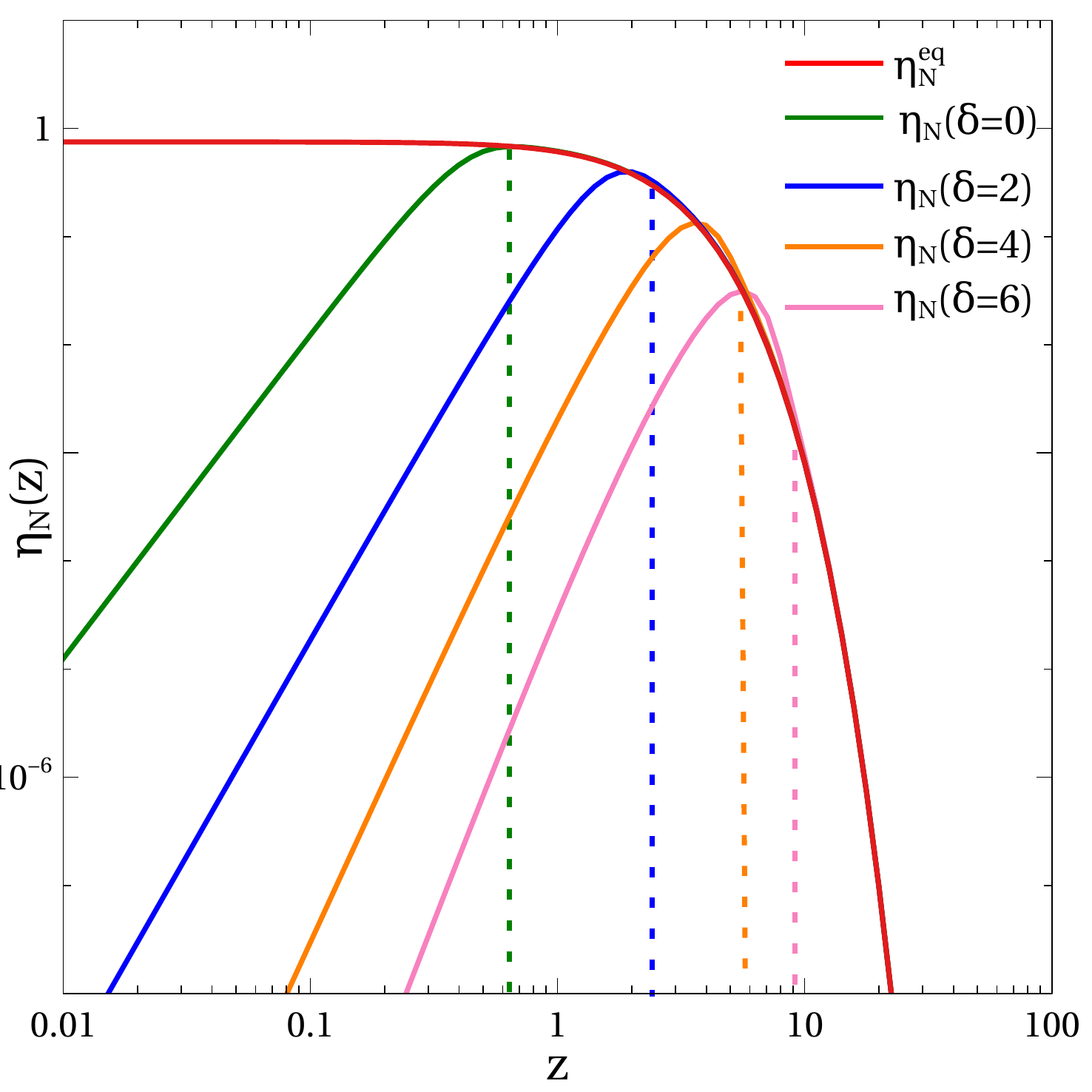}
\includegraphics[width=6cm,height=5cm,angle=0]{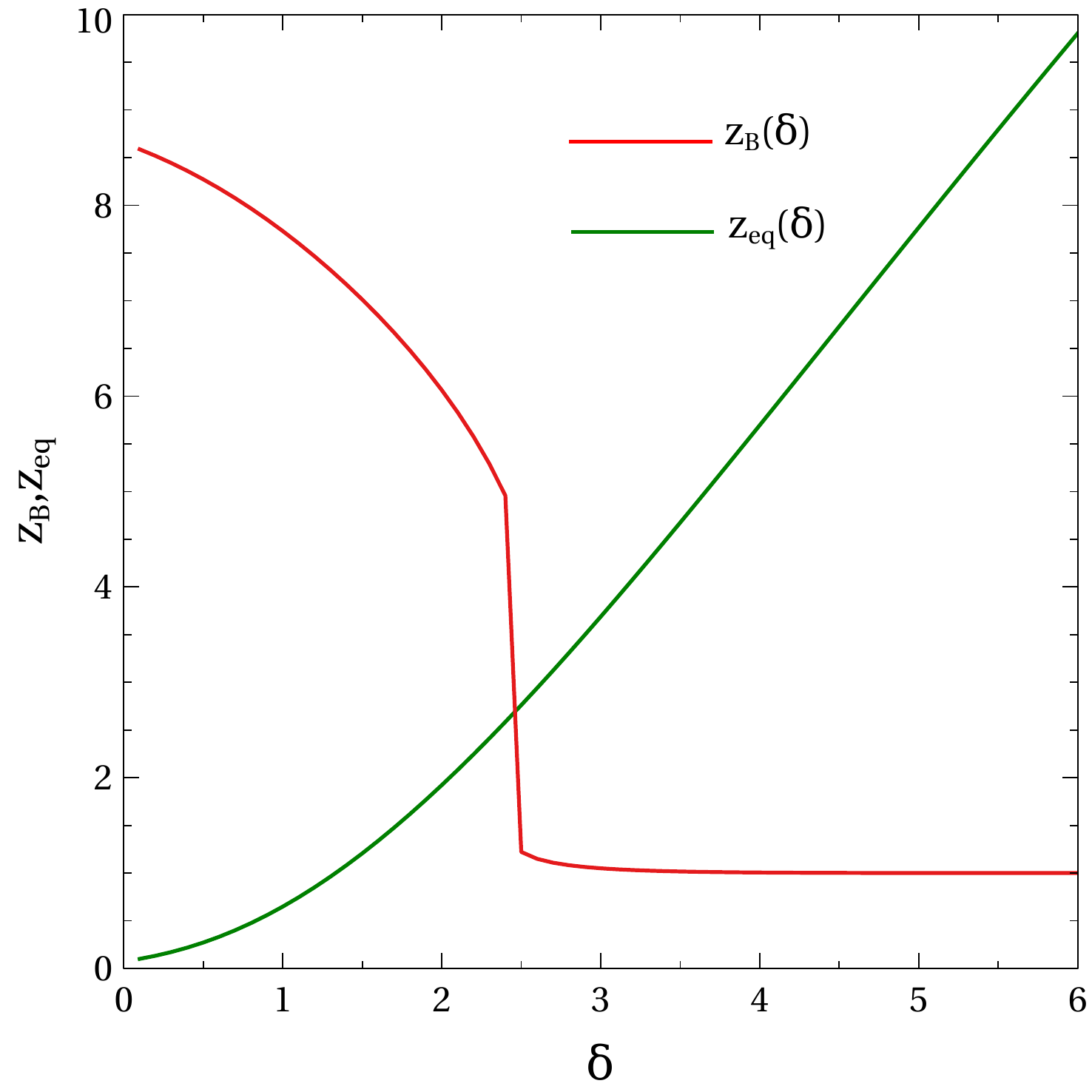}
\caption{{\bf Left:} Plot of $N_1$ abundance (scaled by photon density) for different values of $\delta$. The value of $z$ where the vertical dotted lines touch the abscissa represents $z_{eq}$ for the corresponding curve. {\bf Right:} Variation of $z_{eq}$ and $z_B$ as a function of $\delta$. In both of the figures (left and right) $\gamma$ and $K$ have been kept fixed at $50$ and $100$ respectively. }
\label{zb_zeq} 
\end{center}
\end{figure}
Let us now concentrate on the figure of the right panel. It clearly shows that $z_{eq}$ increases monotonically with $\delta$. The position of maximum contribution to the integral, i.e, $z_B$ was initially
much greater than $z_{eq}$. Unlike $z_{eq}$, $z_B$ shows a steady decrement with increase in $\delta$. Around $\delta \sim 2.5$ $z_B$ shows a sharp (step function like) fall and finally freezes to unity at larger values of $\delta$. Eventually $z_{eq}$ crosses $z_B$ near $\delta \simeq 2.5$ and afterwards $z_{eq}$ always remains greater than $z_B$.   \\

The derivation of analytical expression for $\kappa$ (presented in Sec.\ref{small_d_soln}) is based on three crucial assumptions: (i) $z_{eq} <<1$, which implies that negative contribution ($\kappa^-$) to efficiency factor is negligible and we may take $\kappa \simeq \kappa^+$, (ii) ${\eta^{eq}_N}(z) {\simeq \eta_N}(z)$ for the whole regime, (iii) $z>z_B > z_{eq}$ independent of $\delta$. The concerned plot (Fig.\ref{zb_zeq}, right panel) vividly exhibits the invalidity of all three approximations in the region of $\delta \gtrsim 2$. Therefore now it is inevitable to look for a refined analytical treatment which takes care of all the above mentioned discrepancies and provide a new analytical formula for $\kappa$ which shows good agreement with that of the numerical solution.\\

The remaining part of this section will be devoted in developing new analytical formulas of the efficiency factor for faster expansion (i.e $\delta \gtrsim 2$). Let us first point out the striking differences with our previous analysis (Sec.\ref{small_d_soln}): (i) Here $z_{eq} \gtrsim 1$, which dictates that the negative contribution to the efficiency factor is no longer negligible and the efficiency factor ($\kappa$) should be computed by taking into account the negative ($\kappa^-$) contribution along with the positive ($\kappa^+$) contribution, (ii)the equality of ${\eta^{eq}_N}(z)$ and ${\eta_N}(z)$ holds strictly for $z>z_{eq}$, i.e we can use this equality while calculating $\kappa^+(z>z_{eq})$ , (iii)the calculation of $\kappa^+$ should be carried out through 
straightforward evaluation of the double integral (eq.\ref{kz_mod}) keeping aside the concept of $z_B$ and its associated assumptions since in this case $z_B(<z_{eq})$ lies outside the range of the integration.

\subsubsection{Analytical expression for $\kappa^+(z>z_{eq})$} \label{kp_pos_big_d1}
In this region of $z$, $N_1$ abundance has already reached equilibrium and we can safely replace $\eta_N(z)$
by its equilibrium counterpart which again implies that their derivatives are also equal. This simple assumption allows us to start the calculation of $\kappa^+$ from the following equation
\begin{equation}
\kappa^+(z) = -\frac{4}{3}\int \limits^{z }_{z_{eq}} d z^{\prime} \frac{d {\eta^{eq}_N}(z^\prime)}{d z^{\prime}} ~ e^{- \int \limits^{z }_{z^\prime} W^{\prime}_{\rm ID}(z^{\prime\prime}) dz^{\prime\prime}} ~.
\label{kp_new}
\end{equation}
Using exact analytical expression of equilibrium abundance $(\eta^{eq}_N(z)=(3/8)z^2 \mathcal{K}_2(z))$, its derivative can be easily calculated\footnote{Here we have used the property\cite{arfken} of Bessel function: 
$\frac{\partial}{\partial z}\left( z^n {\mathcal{ K}_n}(z) \right) = -z^n {\mathcal{K}_{n-1}}(z) $ } as 
\begin{equation}
 \frac{d {\eta^{eq}_N}(z)}{d z} = -\frac{3}{8} z^2 \mathcal{K}_1(z)~. \label{dneq_dz}
\end{equation}
Again the nearly accurate analytical fit (valid in the region of large as well as small $z$ ) for the modified Bessel function of first kind is given by\cite{Buchmuller:2004nz}
\begin{equation}
 \mathcal{K}_1(z) \simeq \frac{1}{z}\sqrt{1+ \frac{\pi}{2}z} e^{-z}~.
\end{equation}
In the above integral under consideration the integration variable $z$ is always greater than $z_{eq}$. In the present section we are concerned only about higher values of $\delta$ for which $z_{eq} \gtrsim 4$ and thus the integration variable runs within the interval $4<z<z_f$ ($z_f$ is the upper limit of the integral).       
In the above mentioned interval $\mathcal{K}_1(z)$ can be approximated as 
\begin{equation}
\mathcal{K}_1(z) \simeq \frac{1}{z}\sqrt{\frac{\pi}{2}z} e^{-z} = \sqrt{\frac{\pi}{2}} z^{-1/2} e^{-z}~.
\end{equation}
Using this approximation in eq.\ref{dneq_dz}, expression of $\kappa^+$ (eq.\ref{kp_new}) gets simplified further as
\begin{equation}
 \kappa^+(z) = \frac{1}{2} \sqrt{\frac{\pi}{2}} \int \limits^{z }_{z_{eq}} d z^{\prime} {z^\prime}^{3/2} e^{-z^\prime} ~ e^{- \int \limits^{z }_{z^\prime} W^{\prime}_{\rm ID}(z^{\prime\prime}) dz^{\prime\prime}} ~.
\end{equation}
The above double integration has to be carried out in two steps: integration on $z^{\prime \prime}$, followed by the integration on $z^\prime$ starting from the lower limit of $z_{eq}$ to some arbitrary upper limit $z$.
After evaluation of the integral in the exponent, it will become a function of $(z,z^\prime)$ among which $z$ can be treated as a constant since it is actually the upper limit of the integration. So in the second step the whole integrand becomes function of $z^\prime$. It is then straightforward to perform the integration within the prescribed limit.\\

Upon evaluation of the integral on $z^{\prime \prime}$, the result can be expressed\footnote{Detailed calculation is given in appendix\ref{wid_int1}} as difference of incomplete Gamma functions at points $z^\prime$ and $z$ respectively, i.e
\begin{equation}
 \int \limits^{z }_{z^\prime} W^{\prime}_{\rm ID}(z^{\prime\prime}) dz^{\prime\prime} = 
 C \left[ \Gamma(a,z^\prime) -\Gamma(a,z) \right]~,
\end{equation}
where $C$ and $a$ are constants given by $C=\frac{K}{4}\sqrt{\frac{\pi}{2}} \gamma^{-\delta/2}$, $a=\frac{7+\delta}{2}$. Thus we move a step forward in calculating the efficiency factor $(\kappa^+)$, where the integrand can be expressed as a function of $z^\prime$ only as
\begin{equation}
 \kappa^+(z) = \frac{1}{2} \sqrt{\frac{\pi}{2}} e^{C\Gamma(a,z)}
 \int \limits^{z }_{z_{eq}} d z^{\prime} {z^\prime}^{3/2} e^{-z^\prime} e^{-C\Gamma(a,z^\prime)} ~ .
 \label{kp_new1}
\end{equation}
Analytical evaluation of the above integral requires exact functional form of the incomplete\cite{arfken} Gamma 
function\footnote{Whenever we quote `incomplete Gamma function' it is implicitly understood to be a `upper incomplete Gamma function'} which is in practice expressed in integral representation as 
\begin{equation}
 \Gamma(a,x) = \int \limits^{ \infty }_{ x>1 } e^{-t}~ t^{a-1} dt~, \label{incomp_gamma}
\end{equation}
where $a$ is any non negative real number.
It can also be expressed as an infinite power series in $x$. It is evident that, practically it is impossible to perform the integration using the complete power series. To derive an acceptable analytical expression of $\kappa^+$ without compromising its accuracy, we make some realistic assumptions fit for the situation under consideration. Since we are mainly concerned about high $z$ ($z \gtrsim 4$) region, it would be economical to use the asymptotic form of the above mentioned infinite series, given by\cite{dlmf-nist} 
\begin{equation}
 \Gamma(a,x) \simeq e^{-x} x^{a-1} \sum^{n-1}_{k=0} \frac{u_k}{x^k}~,\label{gamma_asym}
\end{equation}
where the upper limit $n$ of the summation can be taken to be equal to $a$. The numerical factor $u_k$ is actually a product sum, expressed as $u_k =\prod\limits^k_{i=1}  (a-i)$. We have verified that numerical value of $\Gamma(a,x)$ calculated using this above expression is in agreement with its actual value.
In our case the dummy variable $z^\prime >> 1$, which implies that in the asymptotic expansion of 
$\Gamma$ (eq.\ref{gamma_asym})  each term of under the summation is smaller than its predecessor. Thus we need not compute the summation strictly up to the upper limit, instead the series can be truncated at much lower value of $k$. We have checked that for parameter range of our interest (i.e, $x=z_{eq} ~-~z$, $\delta=3~-~10$, $K \geq 10$, $\gamma \geq 40$) it is sufficient to take only the first two terms of the said series, i.e
\begin{equation}
 \Gamma(a,x) \simeq e^{-x} x^{a-1} \left( u_0 + x^{-1}u_1 \right)~. \label{gamma_approx}
\end{equation}
Now we have to expand $\exp[-C\Gamma(a,x)]$ in powers of $\Gamma(a,x)$ using its approximated form (eq.\ref{gamma_approx}) as
\begin{equation}
\exp[-C\Gamma(a,x)]  = \sum\limits^{\infty}_{n=0} \frac{(-C\Gamma(a,x))^n}{n!} =\sum\limits^{\infty}_{n=0} t_n~~.
\end{equation}
Again it can be shown 
that magnitude of successive terms in the above series steadily decreases for our chosen set of parameters, i.e $|t_{n+1}/t_n| <1$. Our purpose can be served by taking into account only the first $(t_0)$ and second $(t_1)$ term of the series, i.e
\begin{equation}
\exp[-C\Gamma(a,x)]  \simeq 1- C\Gamma(a,x)~. \label{exp_approx}
\end{equation}
Incorporating these two (eq.\ref{gamma_approx} and eq.\ref{exp_approx}) approximations in eq.\ref{kp_new1} the integral for evaluation of $\kappa^+$ now becomes
\begin{equation}
\kappa^+(z) = \frac{1}{2} \sqrt{\frac{\pi}{2}} e^{C\Gamma(a,z)}
 \int \limits^{z }_{z_{eq}} d z^{\prime} {z^\prime}^{3/2} e^{-z^\prime} 
 \left[ 1-C e^{-z^\prime} {z^\prime}^{a-1} ( u_0 + {z^\prime}^{-1}u_1) \right]~ . 
\end{equation}
A closer observation of the above integration reveals that it can be expressed as a sum of three integrals given by
\begin{eqnarray}
&& I_1(z) = \int \limits^{z }_{z_{eq}} e^{-z^\prime} {z^\prime}^{3/2}  d z^{\prime} =
\int \limits^{z }_{z_{eq}} e^{-z^\prime} {z^\prime}^{5/2 -1 }  d z^{\prime}~, \\
&& I_2(z) = -Cu_0 \int \limits^{z }_{z_{eq}} e^{-2z^\prime} {z^\prime}^{a+1/2}   d z^{\prime} =
-Cu_0 \left(\frac{1}{2} \right)^{a+3/2} \int \limits^{2z }_{2z_{eq}} e^{-z^{\prime\prime}} {z^{\prime\prime}}^{(a+3/2)-1}   d z^{\prime\prime}~, \\
&& I_3(z) = -Cu_1 \int \limits^{z }_{z_{eq}} e^{-2z^\prime} {z^\prime}^{a-1/2}   d z^{\prime} =
-Cu_1 \left(\frac{1}{2} \right)^{a+1/2} \int \limits^{2z }_{2z_{eq}} e^{-z^{\prime\prime}} {z^{\prime\prime}}^{(a+1/2)-1}   d z^{\prime\prime}~.
\end{eqnarray}
Integrands of the above three equations are identical to that of integral representation of well known Gamma function. The nonzero lower limit and finite upper limit indicate that the result of the integration should be difference of two incomplete Gamma functions evaluated at two distinct points. Thus the functional form of the positive efficiency factor $\kappa^+$ comes out to be
\begin{eqnarray}
 \kappa^+(z) & = & \frac{1}{2} \sqrt{\frac{\pi}{2}} e^{C\Gamma(a,z)} \left[ I_1(z) + I_2(z) +I_3(z) \right]~,\label{kp_pos_big_d} 
\end{eqnarray}
where
\begin{eqnarray}
&& I_1(z) = \Gamma(5/2,z_{eq}) -\Gamma(5/2,z) ~,\label{I1} \\ 
&& I_2(z) = - Cu_0 \left(\frac{1}{2} \right)^{a+3/2}\left \{ \Gamma(a+3/2,2z_{eq}) -\Gamma(a+3/2,2z) \right\}~, \label{I2}\\
&& I_3(z) = - Cu_1 \left(\frac{1}{2} \right)^{a+1/2}\left \{ \Gamma(a+1/2,2z_{eq}) -\Gamma(a+1/2,2z) \right\}~. \label{I3}
\end{eqnarray}

\subsubsection{Analytical expression for $\kappa^-(z<z_{eq})$} \label{kp_neg_big_d1}
Presently we are concerned with the cases of larger values of $\delta$, for which RH neutrinos take longer time to reach equilibrium, i.e $z_{eq}>1$. Although in this section we are considering strong washout, the bigger values of $\delta$ drives $z_{eq}$ towards higher end. The analytic expression for $\kappa^-$ should be the same
as that of eq.\ref{kp_neg} which has been presented in Sec.\ref{kp_neg_small_d}.
Using the same argument as Sec.\ref{kp_neg_small_d} it is not difficult to understand that this expression of $\kappa^-$ is valid within the interval $z_{in}<z<z_{eq}$, beyond which it may be safely neglected without affecting the final result.

\subsubsection{Analytical expression for total efficiency factor $\kappa(z)$}\label{kp_tot_big_d}
The total efficiency factor is represented as the sum of positive and negative contribution where these contributions are active in the intervals $(z_{eq}<z<\infty)$ and $(z_{in}<z<z_{eq})$ respectively. It can be expressed in a compact form as
\begin{equation}
 \kappa (z)  \simeq  \kappa^+(z) \Theta (z-z_{eq})  + \kappa^-(z) \Theta (z_{eq} -z) ~.
\end{equation}
Replacing the actual analytical expressions of $\kappa^+$ (eq.\ref{kp_pos_big_d}) and $\kappa^-$ (eq.\ref{kp_neg}) the final form of $\kappa$ turns out to be
\begin{equation}
 \kappa(z) = \frac{1}{2} \sqrt{\frac{\pi}{2}} e^{C\Gamma(a,z)} \left[ I_1(z) + I_2(z) +I_3(z) \right] \Theta (z-z_{eq}) -2 \left( 1 - e^{-\frac{2}{3}\eta_N(z)} \right) \Theta (z_{eq} -z) ~,\label{kp_total_big_d}
\end{equation}
where the proper functional from of $I_1(z),I_2(z),I_3(z)$ can be found from eqs.(\ref{I1}-\ref{I3}) and the RH neutrino abundance $\eta_N(z)$ is given in eq.\ref{mod_N}. We now examine the asymptotic behaviour of the efficiency factor. The final value of $B-L$ asymmetry is directly proportional to the final efficiency factor (denoted by $\kappa_f$) which is actually the value $\kappa$ at some high $z$ value where the asymmetry practically gets frozen. $\kappa_f$ can be computed from eq.\ref{kp_total_big_d} by taking the limit $z \rightarrow \infty$. In the said limit the negative contribution can be easily omitted, leaving only the positive part given by
\begin{equation}
\kappa_f = \kappa (z \rightarrow \infty) = \frac{1}{2} \sqrt{\frac{\pi}{2}} e^{C\Gamma(a,z\rightarrow \infty)} 
\left[ I_1(z\rightarrow \infty) + I_2(z \rightarrow \infty) +I_3(z\rightarrow \infty) \right]
\end{equation}
In this case expressions for $I_1(z),I_2(z),I_3(z)$ becomes far more simplified since the corresponding upper incomplete Gamma functions will vanish in the large $z$ limit. We now impose the $z \rightarrow \infty$ limit in eqs.(\ref{I1}-\ref{I3}) and thereby after a few steps of simplification we arrive at the expression of final efficiency factor as
\begin{eqnarray}
\kappa_f  = & \frac{1}{2}  & \sqrt{\frac{\pi}{2}}  \left[  
\left\{ \Gamma(5/2,z_{eq})  \right\}
-Cu_0 \left(\frac{1}{2} \right)^{a+3/2}\left \{ \Gamma(a+3/2,2z_{eq})  \right\} \right. \nonumber \\
 & - & \left. Cu_1 \left(\frac{1}{2} \right)^{a+1/2}\left \{ \Gamma(a+1/2,2z_{eq})  \right\}  
\right]~. \label{kpf_total_big_d}
\end{eqnarray}
\subsubsection{Reliability of the refined analytical formulas for faster expansion}
Although the previously derived (in Sec.\ref{kp_total_small_d}) simple analytical formulas of $\kappa$ efficiently reproduces the correct numerical value for nonzero but small $\delta$, it fails badly as 
$\delta$ increases beyond a certain limit, as shown in Sec.\ref{plots_small_d}. 
To circumvent this problem we have derived new analytical formulas for the efficiency factor valid for faster expansion (i.e larger values of $\delta$, more specifically for $\delta \gtrsim 2$) in the preceding subsections. \\

\begin{figure}[h!]
\begin{center}
\includegraphics[width=5.5cm,height=5cm,angle=0]{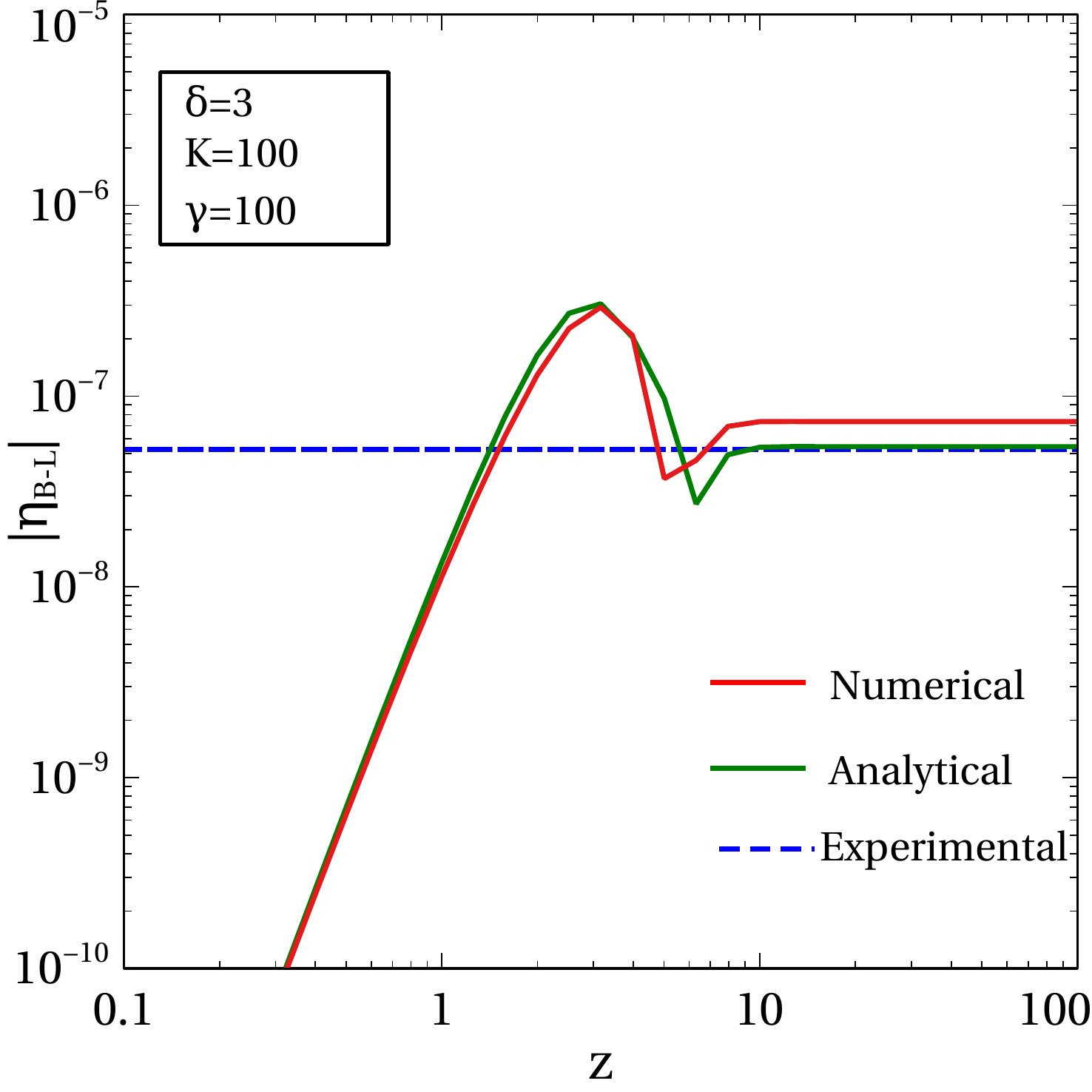}
\hspace{.5cm}
\includegraphics[width=5.5cm,height=5cm,angle=0]{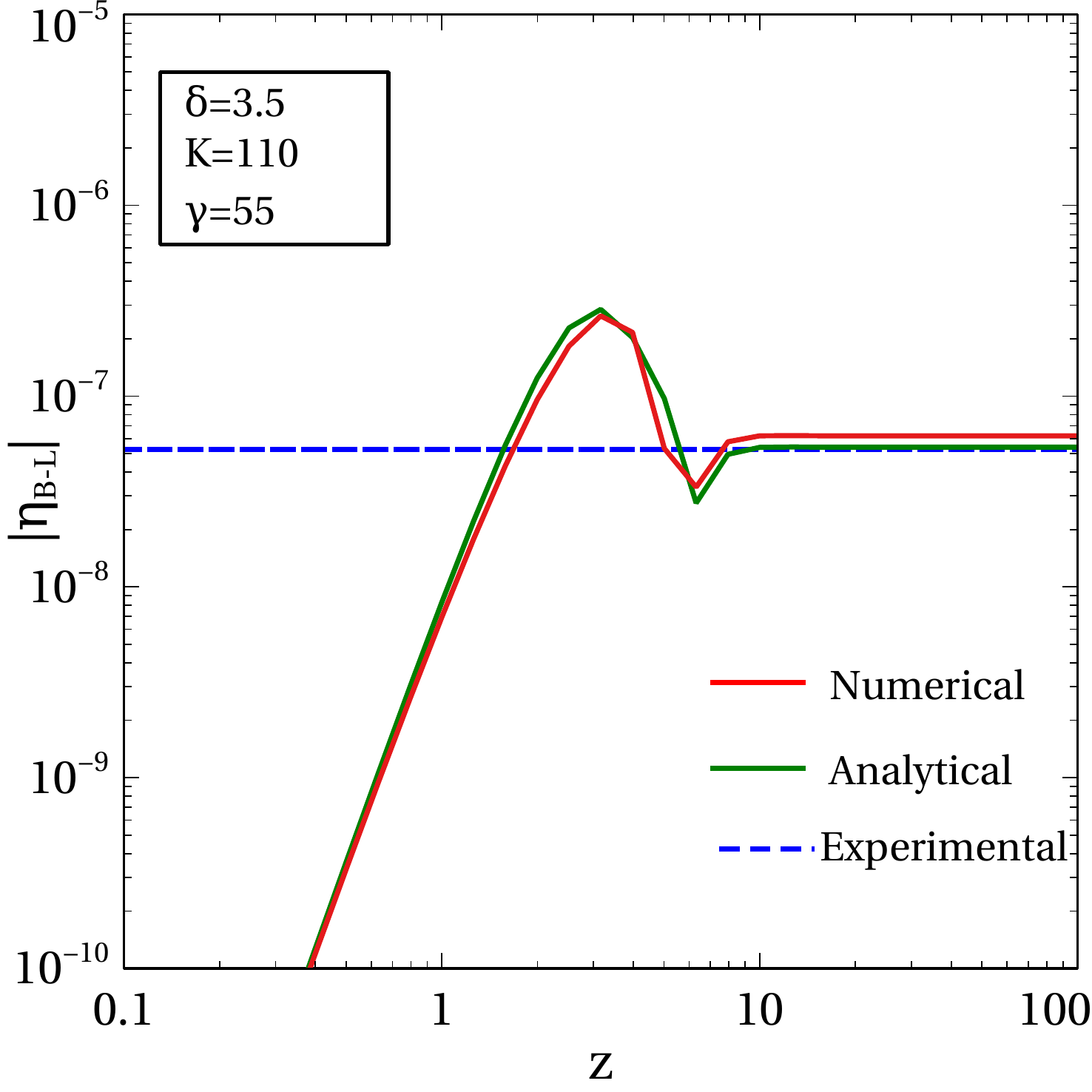}
\caption{Evolution of $\eta_{B-L}(z)$ estimated through numerical and analytical solution of Boltzmann equations for $\delta=3$(left) and $\delta=3.5$(right). A fixed value of the CP asymmetry parameter $\epsilon_1=-2\times10^{-6}$ has been used for all the cases. 
Using the experimental bound(eq.\ref{expt-asym}) on $\eta_B$ in eq.\ref{conv}, the final value of $B-L$ asymmetry can be constrained as $5.24\times10^{-8}<|\eta_{B-L}(z\rightarrow\infty)|<5.38\times10^{-8}$ ($95\%$ CL). The allowed range of 
$\eta_{B-L}$ is represented by the horizontal blue dashed line.
Analytically calculated (using eq.\ref{kp_total_big_d}) asymmetry ($\eta_{B-L}(z)$) shows fair agreement with that of the numerical solution. The values of the chosen set of parameters $(\delta,K,\gamma)$ are indicated inside the box in the respective plots.}
\label{del3_4}
\end{center}
\end{figure}

\begin{figure}[h!]
\begin{center}
 \includegraphics[width=5.5cm,height=5cm,angle=0]{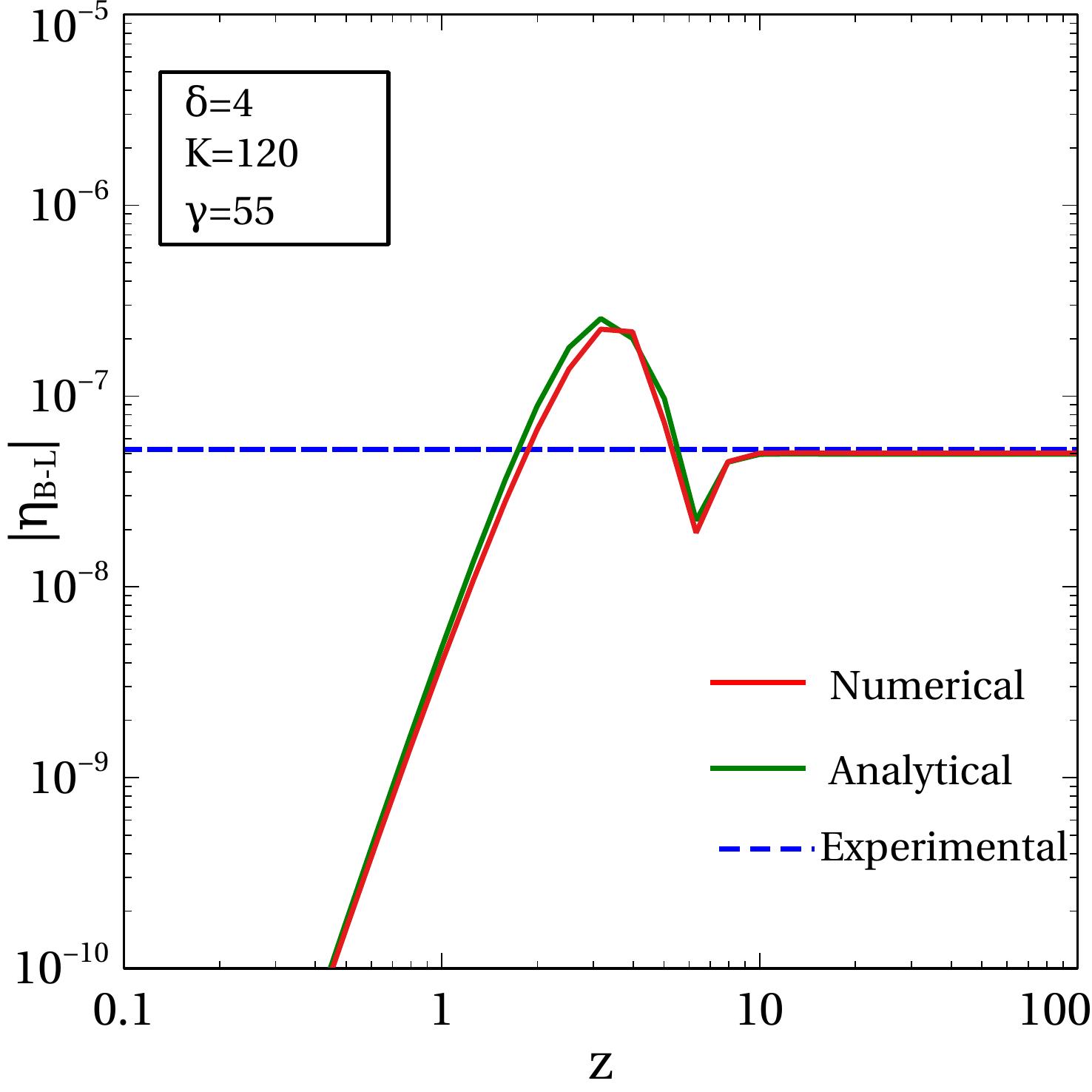}
\hspace{.5cm}
\includegraphics[width=5.5cm,height=5cm,angle=0]{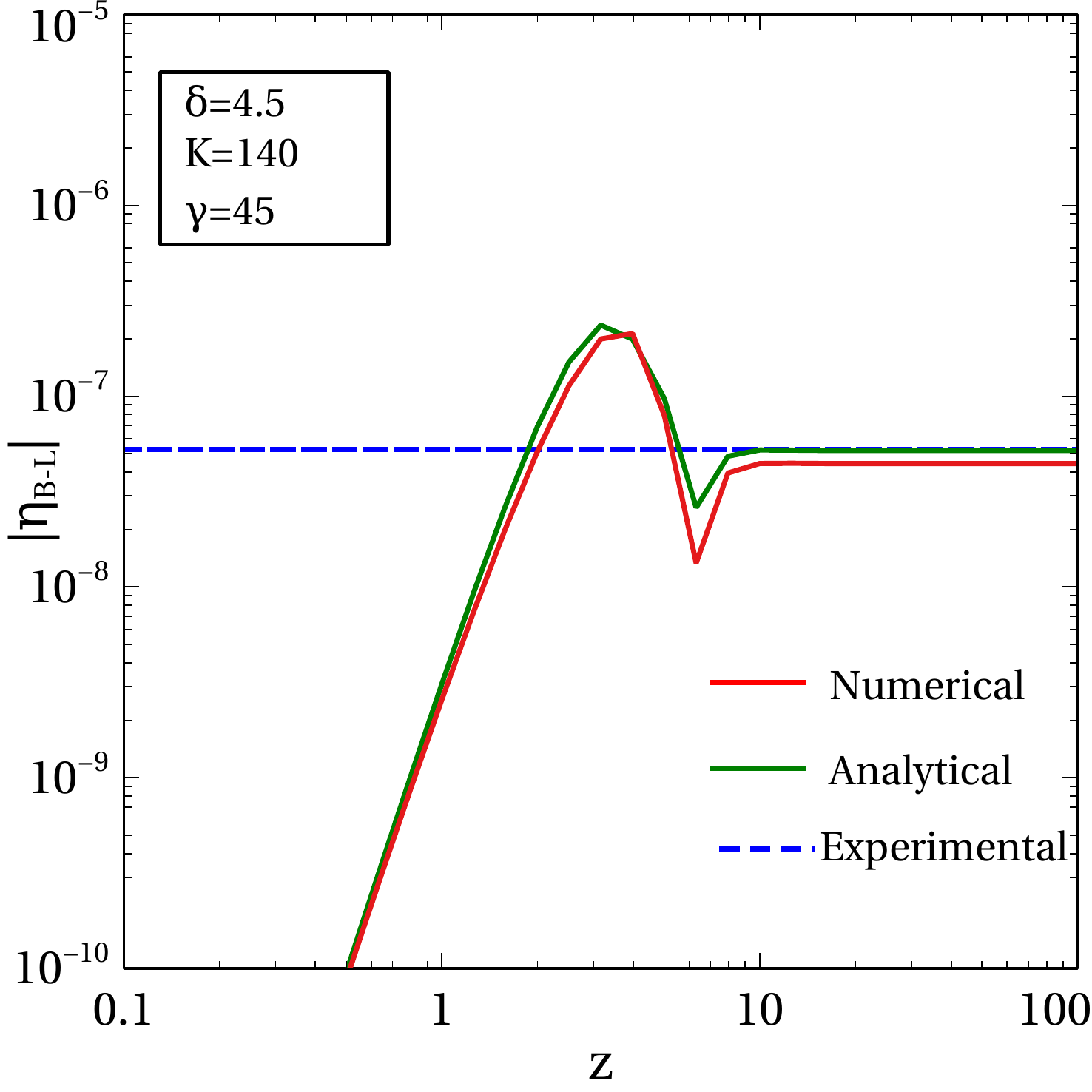}
\caption{Evolution of $\eta_{B-L}(z)$ estimated through numerical and analytical solution of Boltzmann equations for $\delta=4$(left) and $\delta=4.5$(right). }
\label{del5_6}
\end{center}
\end{figure}

Again the accuracy of these new analytical formulas (eq.\ref{kp_total_big_d}, eq.\ref{kpf_total_big_d}) have to be tested by comparing the numerical value of $\kappa(z)$ calculated using these analytical formulas with that of the actual numerical solution of Boltzmann equations. We compute $\kappa$ with the said analytical and numerical procedure and plot (Fig.\ref{del3_4}, Fig.\ref{del5_6}) the corresponding $B-L$ asymmetries as a function of $z$ in the same figure. This exercise is repeated for four different values of $\delta$, $(\delta=3,3.5,4,4.5)$ for which the previous simplified formula (eq.\ref{kp_tot}) ceases to hold.
Before showing the plots, a few remarks about the final value of the $B-L$ asymmetry are in order.
Since we are not working with any specific flavour symmetric model, we have ample scope of tuning the available free parameters in order to get the final asymmetry within the range allowed by the experiments. Thus we have judiciously chosen a few specific combination of parameters such that the final asymmetry matches with that of the experimental value. Corresponding graphical representation of evolution of the asymmetry parameter $(\eta_{B-L}(z))$ are shown in Fig.(\ref{del3_4},\ref{del5_6}). 
The plots clearly show that the refined analytical formulas ((eq.\ref{kp_total_big_d}, eq.\ref{kpf_total_big_d})) for the efficiency factor (both $\kappa(z)$ and $\kappa_f$ )  matches very well with that of the numerical solution. \\

Although we have not been able to derive an unique analytical expression for the efficiency factor $\kappa$ suitable for 
all values of $\delta$,
we are successful in finding out two separate expressions 
for $\kappa$ (eq.\ref{kp_tot}) one valid only for smaller\footnote{To be specific `smaller' (`larger') implies $\delta \lesssim 2$ 
($\delta \gtrsim 2$). }  values $\delta$ whereas the other (eq.\ref{kp_total_big_d}) efficiently reproduces correct value of asymmetry for relatively larger values of $\delta$. 
These formulas will be quite helpful in the problems where we need to scan the whole multidimensional parameter space to check whether there exists any common parameter space satisfying both oscillation phenomenology and baryon asymmetry bound. In these cases generally it becomes inevitable to solve the set of Boltzmann equations for each set of points in the parameter space. Sometimes the time consumed in this task is so huge that it becomes practically impossible to tackle the problem through this approach. These newly introduced analytical formulas for efficiency factor will come in handy in those situations. The scan of whole parameter space can be completed within a feasible time with the help of these formulas.

\section{Effects of deviation from standard cosmology on the existing bounds} \label{deviation-bounds}
Let us first count the number of independent parameters needed to describe the full Type-I seesaw theory and examine whether they can be constrained by the available neutrino oscillation data. Although the Type-I seesaw (with three RH neutrinos) is the simplest among its all variants, it requires $21$ independent parameters\cite{Strumia:2006db}, whereas the number of experimental observables is limited to $12$ ($3$ charge lepton masses, $3$ light neutrino masses, $3$ leptonic mixing angles, $1$ Dirac CP phase, $2$ Majorana phases). Therefore it is evident that full Type-I seesaw theory can not be constrained only by low energy neutrino oscillation data. However it is possible to impose a lower bound on the lightest RH neutrino $(N_1)$ mass, if it is assumed that the entire matter-antimatter symmetry is produced by decay of $N_1$. It has been shown in\cite{Davidson:2002qv} that the theoretical upper bound on the $N_1$ decay generated asymmetry can be expressed as
\begin{equation}
\epsilon_1 \lesssim \frac{3 M_1}{ 16 \pi v^2} \left( m_3 - m_1 \right) ~, \label{lim_eps}
\end{equation}
where $M_1$ is the mass of lightest RH neutrino, $m_i(i=1,2,3)$ denotes light neutrino mass eigenvalue and $v$ stands for SM VEV. In case of normally ordered light neutrino mass spectrum, we have $m_3 \gg m_2 > m_1$.
Therefore the atmospheric mass squared difference can be approximated as $\sqrt{\Delta m^2_{\rm atm}}=\sqrt{m^2_3 -m^2_1} \simeq m_3$. Now the said upper limit on CP asymmetry becomes
\begin{equation}
 \epsilon_1 \lesssim \frac{3 M_1}{ 8 \pi v^2}\sqrt{\Delta m^2_{\rm atm}}  ~. \label{lim_eps1}
\end{equation}
Similarly the final baryon asymmetry produced due to this CP asymmetry should also have an upper bound given by
\begin{equation}
 \eta_B \leq d \epsilon_1 \kappa_f ~,\label{eta_B1}
\end{equation}
where $\kappa_f$ is the final efficiency factor and $d$ contains the sphaleron\cite{tHooft:1976rip,tHooft:1976snw} conversion factor $(\alpha_{ sph})$ along with the dilution\cite{Buchmuller:2004nz} factor $(F=N^{\rm rec}_\gamma/N^\ast_\gamma)$ as $d= \alpha_{ sph}/F \simeq 1.2 \times 10^{-2}$. After expressing $\epsilon_1$ in terms of $M_1$ in eq.\ref{eta_B1} the baryon asymmetry bound shows explicit dependence on $M_1$ as 
\begin{equation}
\eta_B \leq 1.2 \times 10^{-2}  \frac{3 M_1}{ 8 \pi v^2}\sqrt{\Delta m^2_{\rm atm}} \kappa_f~,
\end{equation}
which can be inverted to get the lower bound on $M_1$ as
\begin{equation}
M_1 \geq \frac{1}{1.2 \times 10^{-2}  } \frac{8 \pi v^2}{3} \frac{\eta^{\rm obs}_B}{\sqrt{\Delta m^2_{\rm atm}}} 
\frac{1}{\kappa_f}~. \label{bound_M11}
\end{equation}
If the washout effect is neglected (i.e $\kappa_f \simeq 1$), the lowest mass of $N_1$ required to produce the observed baryon asymmetry $(\eta^{\rm obs}_B \times10^{10}\simeq 6.29 ~-~6.46 )$\cite{Planck:2018vyg,Planck:2013pxb,Aghanim:2016yuo} comes out to be $M_1 \gtrsim 10^9$ GeV.\\

We already know a few loop holes through which this bound can be relaxed. Quasi-degenerate RH neutrinos of few TeV mass can give rise to resonantly\cite{Pilaftsis:1997jf,Pilaftsis:2003gt} enhanced CP asymmetry which is enough to produce the observed baryon asymmetry. Inclusion of flavour\cite{Abada:2006fw,Abada:2006ea,Antusch:2006cw,Adhikary:2014qba,Samanta:2018hqm,Chakraborty:2020gqc} effects can also bring down the washout of asymmetry significantly resulting in survival of greater amount of asymmetry compared to that of unflavoured case. All these inferences are drawn based on solution of Boltzmann equations in standard radiation dominated cosmology. So the results and corresponding conclusions are likely to change if we modify the cosmological history of evolution. In what follows we try to figure out whether the effect of faster expansion can alter the final value of the baryon asymmetry which is to be compared with the experimental value. Or in other words is it really possible to get a larger value of final baryon asymmetry in modified cosmology compared to standard one for the same value of lightest RH neutrino mass. 
\subsection{Lower bound on lightest right handed neutrino mass}\label{mod-DI-bound}
The eq.\ref{bound_M11} help us infer that except $\kappa_f$, all the quantities in the R.H.S. are either experimental numbers or some constants. Thus we can express this equation in a concise form as
\begin{equation}
M_0 = \frac{A}{\kappa^0_f}~, \label{bound_M12}
\end{equation}
where $A$ takes care of all the constant factors and experimental numbers\footnote{$A$ does not contain any parameter related to  expansion rate of the Universe}. The `$0$' superscript  
on $\kappa_f$ signifies that the efficiency factor has been calculated assuming standard radiation dominated cosmology and $M_0$ is the corresponding lower limit on the mass of lightest RH neutrino. In the preceding sections of this work we have shown that the efficiency factor indeed changes in modified cosmology and derived
the approximate analytical expression for the same. Let us now try to understand how the lower bound ($M_0$)
can be modified in case of faster (compared to the standard case) expansion. The simple equation (\ref{bound_M12}) guide us to express the modified lower limit ($M$) on $N_1$ mass in terms of ratio (standard to modified) of efficiency factors as
\begin{equation}
M= M_0 \frac{\kappa^0_f}{\kappa_f} ~.
\end{equation}
Using the analytical formula of final efficiency factors (eq.\ref{kpf} for standard cosmology and eq.\ref{kf_mod1} for modified cosmology with low $\delta$) in the limit of very strong washout, the ratio of the lower limits can be expressed as functions of the parameters $K,\delta,\gamma$, i.e
\begin{equation}
\frac{M}{M_0} = \frac{\kappa^0_f}{\kappa_f}= \frac{z_B(K) f(z_B)}{z^0_B(K)}= 
\left( \frac{z_B(K)}{z^0_B(K)} \right) \left[ 1 + \left(\frac{\gamma}{z_B} \right)^\delta \right]^{-1/2} ~.
\end{equation}
It will be easier to perceive the nature of this ratio if we represent it in terms of logarithms, i.e we define
\begin{eqnarray}
\Delta & = & \log \left( \frac{M}{M_0} \right) \label{delta_defn} \\
       &=  & \log \left( \frac{\kappa^0_f}{\kappa_f} \right) \label{ratio_kp_kp0}\\
       & = & \log \left( \frac{z_B(K)}{z^0_B(K)} \right) -\frac{1}{2} \log \left[ 1 + \left(\frac{\gamma}{z_B} \right)^\delta \right]~, \label{ratio_MM0}
\end{eqnarray}
where we have used the approximate analytical formula (eq.\ref{kf_mod1}, appropriate for $\delta \lesssim 2$) for efficiency factor to arrive at eq.\ref{ratio_MM0} from eq.\ref{ratio_kp_kp0}. 
Similar expression for $\Delta$ can also be derived for higher values of $\delta(\gtrsim 2)$ using eq.\ref{kpf} 
and eq.\ref{kpf_total_big_d}. 
However we do not present its explicit form here.\\

\begin{figure}[h!]
\begin{center}
\includegraphics[width=7cm,height=6cm,angle=0]{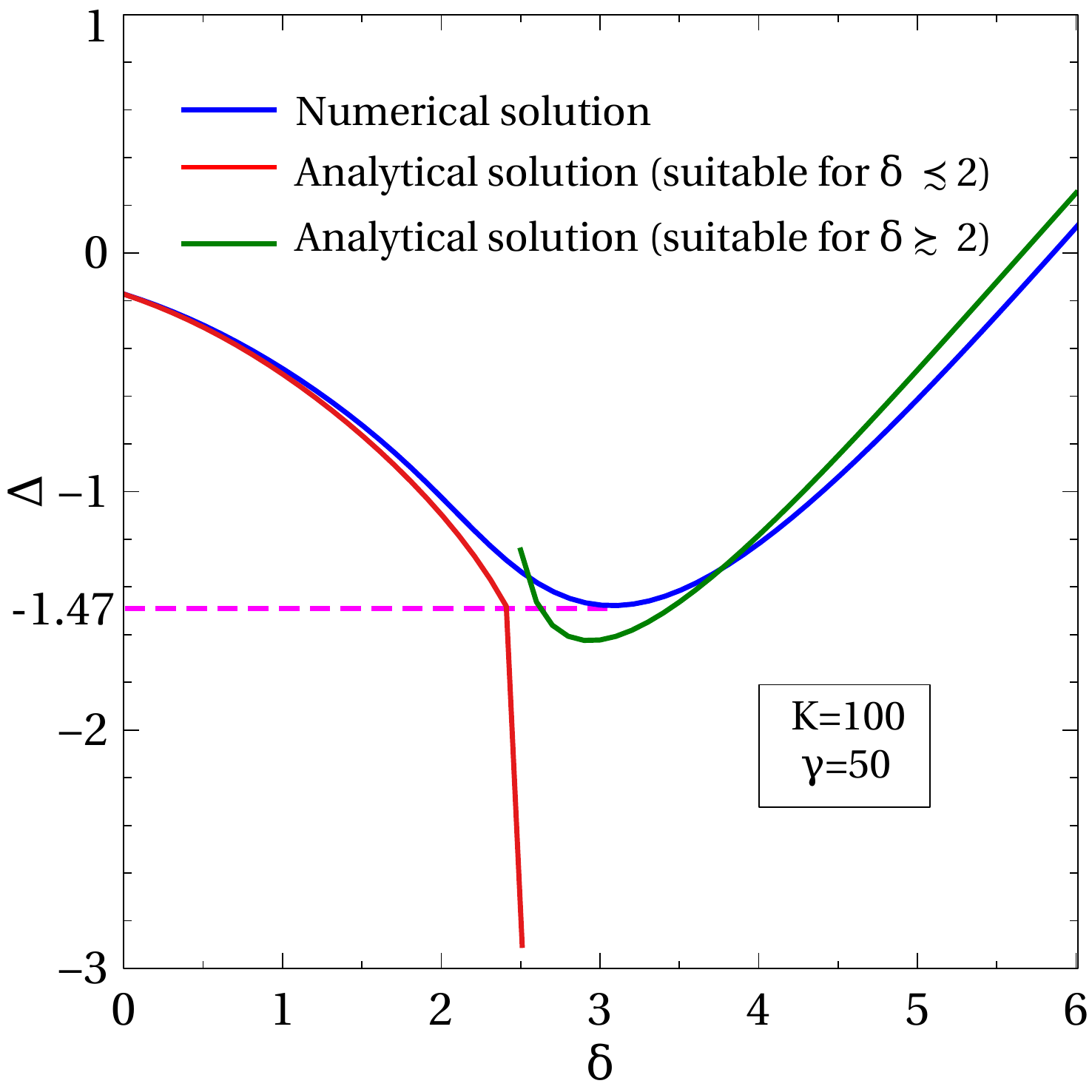}
\hspace{.5cm}
\includegraphics[width=7cm,height=6cm,angle=0]{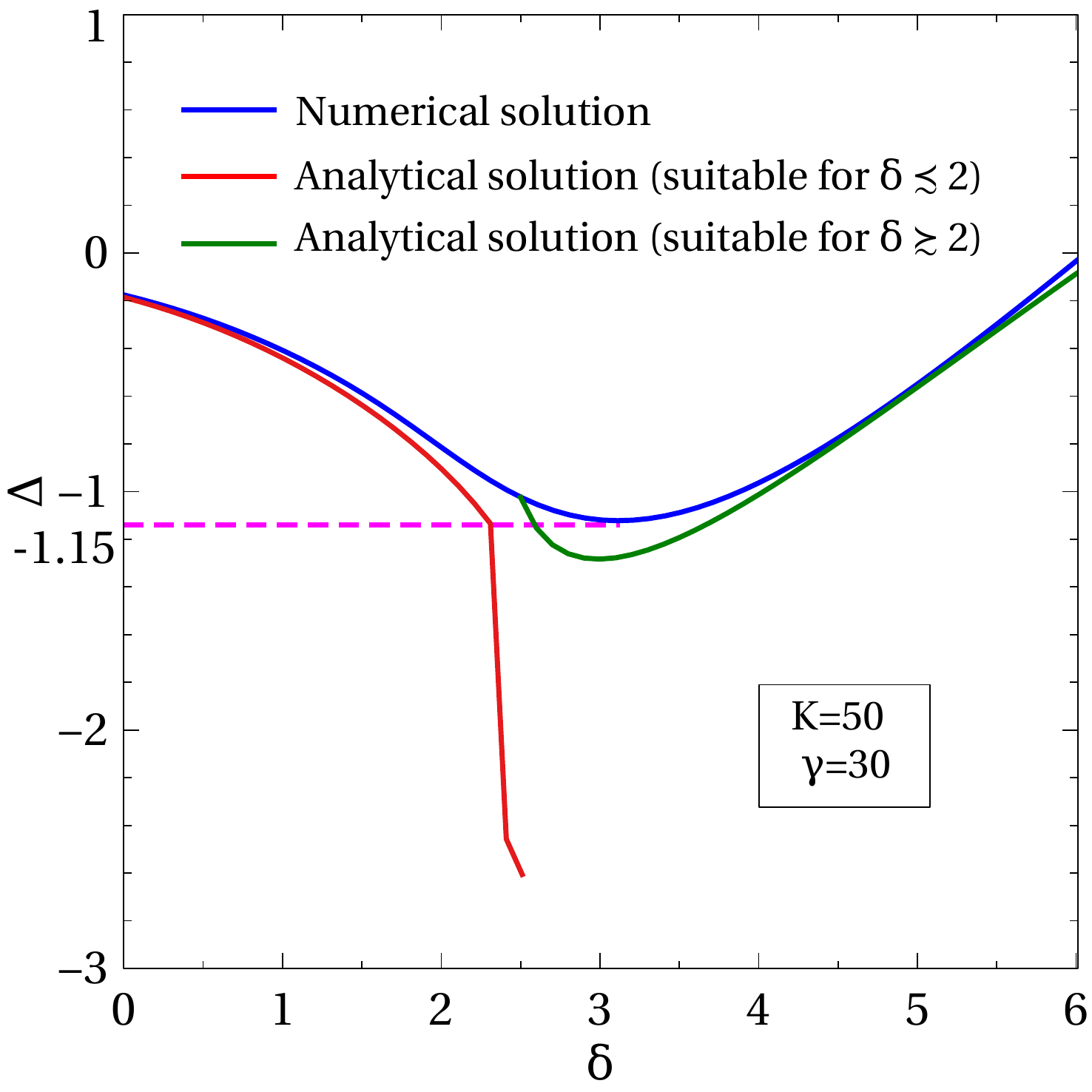}
\caption{Plot of $\Delta$ as a function of $\delta$. The horizontal dashed line is incorporated to mark the minimum achievable value of 
$\Delta$. {\bf left:} ($K=100,\gamma=50$) has been kept fixed throughout. {\bf right:} ($K=50,\gamma=30$) has been kept fixed throughout. }
\label{delta1_delta2}
\end{center}
\end{figure}

In our following analysis, the final efficiency factor $\kappa_f$ has been evaluated
analytically (using approximate expression of $\kappa_f$, i.e eq.\ref{kf_mod1} for $\delta \lesssim 2$ and eq.\ref{kpf_total_big_d} for $\delta \gtrsim 2$) as well as numerically (solving the Boltzmann equation).   
Thus the above mentioned logarithmic ratio $\Delta$ is also computed twice, analytically and numerically.
In Fig.\ref{delta1_delta2}, 
we examine variation of $\Delta$ with $\delta$ while $K$ and $\gamma$ are kept fixed\footnote{It is clear from Fig.\ref{delta1_delta2}, that the two distinct analytical formulas are in fair agreement with the numerical result above and below a critical value of $\delta$. However, both of analytical approximations fails badly near the critical value of $\delta$, which is around $2.5$ }. 
This exercise is repeated for two different fixed values of the set $(K,\gamma)$ and the corresponding variations of $\Delta$ is depicted in left and right panel of the same figure. A negative value of $\Delta$ indicates that $M<M_0$, i.e the lower limit of $N_1$ mass can be brought down further when the effect of faster expansion is taken into account.
It is clear from Fig.\ref{delta1_delta2}, that the logarithmic ratio $\Delta$ is negative for almost the entire range of $\delta$ shown in the figures. The ordinate of the horizontal dashed line in the figure signifies the lowest achievable value of $\Delta$ for a fixed set of 
$(K,\gamma)$. Thus the maximum possible relaxation on the lower bound of $N_1$ mass is $(10^{1.47}\simeq)~30$ times for ($K=100,\gamma=50$) and $(10^{1.15}\simeq)~14$ times for ($K=50,\gamma=30$).   
So it can be concluded that the lower limit on lightest RH neutrino mass required to meet the observed baryon asymmetry bound can be relaxed (or brought down) for a wide region of suitably chosen parameters.\\

Apparently, eq.\ref{ratio_MM0} hints towards a monotonically decreasing nature of $\Delta$ (with increase in $\delta$), or in other words it seems from eq.\ref{ratio_MM0}, that the lower bound on the lightest RH neutrino mass can be relaxed indefinitely by increasing the value of $\delta$. However, the actual nature of $\Delta$ (the continuous blue line in Fig.\ref{delta1_delta2}) clearly contradicts this idea. The said figure shows that, although at the beginning,
$\Delta$ decreases with increase in $\delta$, after a critical value of $\delta$ the nature of $\Delta$ becomes completely opposite and it shows a increasing nature with $\delta$. This behaviour of $\Delta$ can be explained with the help of effective decay parameter $K_{\rm eff}$ (eq.\ref{Keff}) which directly controls the process of production and washout of asymmetry. Now, $K_{\rm eff}$ always decreases with $\delta$. It is to be noted that decrease in $K_{\rm eff}$ not only reduces the washout, but also the asymmetry production (this behaviour is clearly depicted in the left panel of Fig.\ref{zb_zeq}, where we have shown that for $\delta \gtrsim 2$ the $N_1(z)$ abundance takes longer time to equilibrate and $N_1(z)$ touches the equilibrium curve when its downfall has already started). There is a critical value of $\delta$ (for a fixed set of $(\gamma, K)$) for which the reduction of washout surpluses the reduction of asymmetry production and thus giving a positive change\footnote{Increase(decrease) in $\kappa_f$ implies decrease(increase) in $\Delta$, i.e lowering(raising) of the DI bound.} in the efficiency factor. However, above that critical value of delta, reduction of asymmetry production dominates and as a result the efficiency factor decreases.\\

\begin{figure}[h!]
\begin{center}
\includegraphics[width=7cm,height=6cm,angle=0]{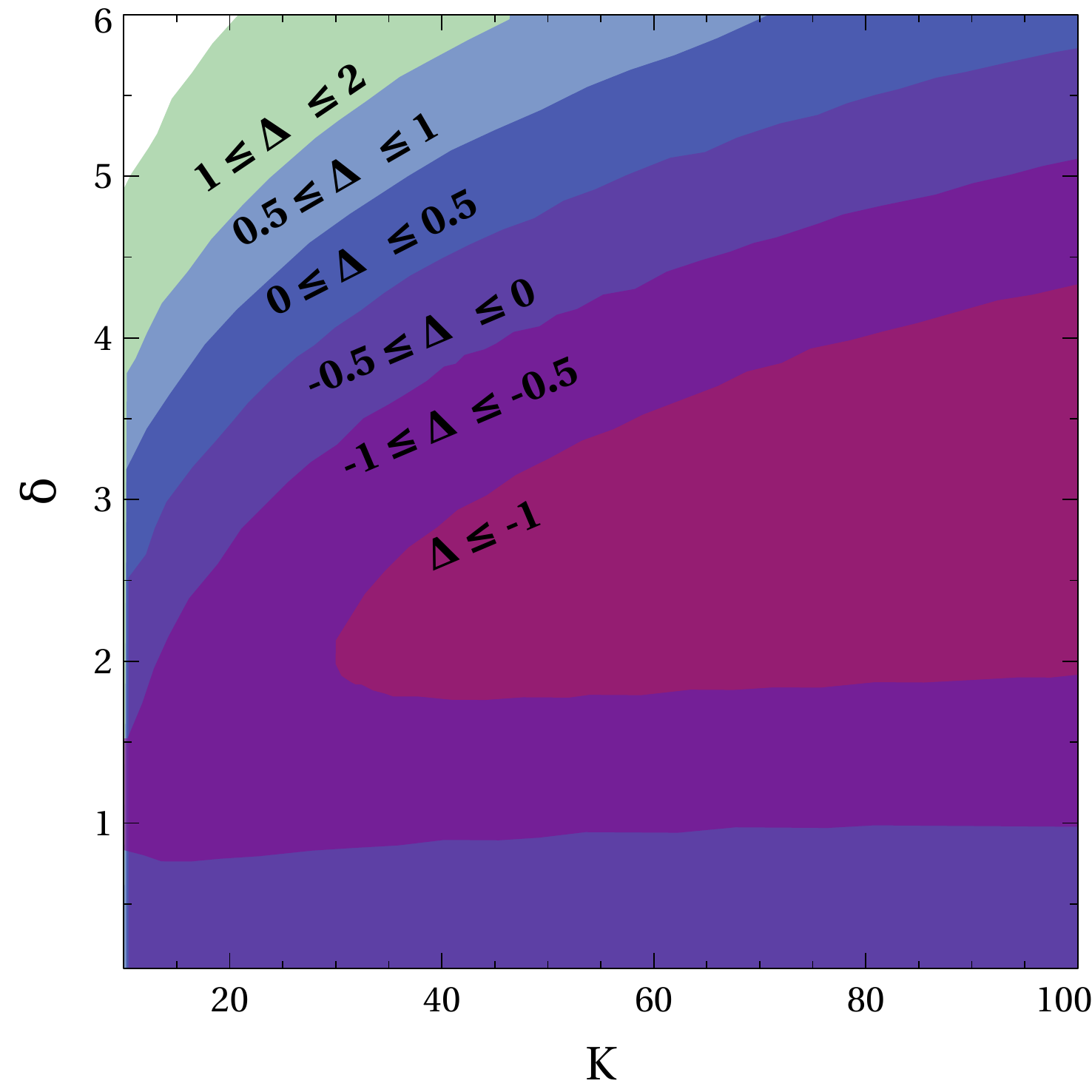}
\hspace{0.5cm}
\includegraphics[width=7cm,height=6cm,angle=0]{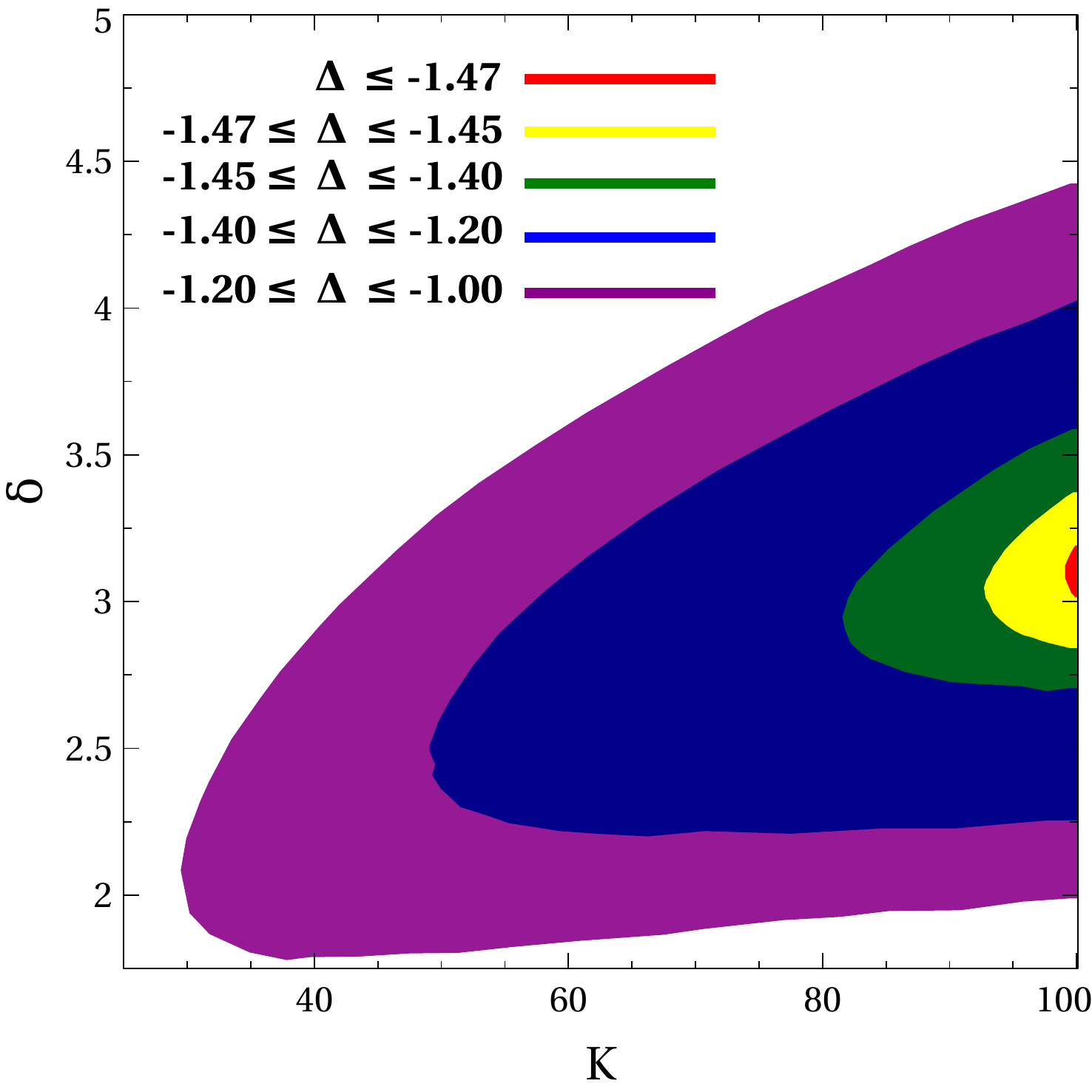}
\caption{{\bf Left:} Contours of logarithmic ratio of lower limits (modified cosmology to standard cosmology) of $N_1$ mass in $K-\delta$ plane derived from baryon asymmetry bound. $\gamma$ is always kept fixed at $50$. {\bf Right:} A detailed view of the 
region $\Delta \leq -1$. The small red patch signifies that lowest possible value of $\Delta$ is $-1.47$ (for $K \leq 100$).}
\label{contour_Delta}
\end{center}
\end{figure}

Although this simple analytical formula (eq.\ref{ratio_MM0}) is quite competent in drawing correct inference 
within a shorter time frame, it does not hold good for larger values of $\delta(\gtrsim 2)$. Thus to get a comprehensive picture of modification of lower bound on $N_1$ mass when the parameters $(K,\delta)$ are varied simultaneously 
we resort to numerical solution of Boltzmann equations for same sets of $(K,\delta)$ once for standard cosmology and then for fast expansion. Solution of Boltzmann equations provide efficiency factors which enable us to calculate the logarithmic ratio $\Delta$ for each combination of $(K,\delta)$. For systematic understanding of the result we draw contours (Fig.\ref{contour_Delta}) of $\Delta$ in $(K,\delta)$ plane. 
The contour plot of the right panel has been introduced to examine, whether more contours (with $\Delta<-1$) can be drawn inside the 
$\Delta=-1$ contour. It has been revealed from the said plot that the minimum achievable value of $\Delta$ is $-1.47$ (shown by the little red patch) if we restrict ourselves in the region $K \leq 100$.  
It is clear from Fig.\ref{contour_Delta} that for the larger portion of the parameter space the logarithmic ratio $\Delta$ 
(eq.\ref{delta_defn}) turns out to be negative, i.e $M<M_0$. We can put it in an other way that for most of the combinations of $(K,\delta)$
the lower bound on $N_1$ mass decreases. It allows us to simply conclude that the introduction of modified cosmology 
further brings down the lower limit of $N_1$ mass, i.e the stringent Davidson-Ibarra bound can be relaxed to some extent when the expansion of the Universe is faster than the standard radiation dominated scenario. \\\\

The discussions and calculations presented so far in this manuscript are valid in strong washout regime. For the sake of completeness we have studied the effect modified cosmology in weak washout regime too. A few simple mathematical calculations assuming the washout effects to be very small suggests that modification of expansion rate of Universe does not incorporate any significant (or noticeable) change in the final results compared to the standard radiation dominated case. Thus we have omitted elaborate discussions of weak washout in the main text. However the said mathematical calculations which enabled us to draw the above conclusion are given in appendix\ref{weak_wash}.

\section{Summary and conclusion}\label{summary}
Analytical solutions of the Boltzmann equations for leptogenesis in standard radiation dominated cosmology already exists in literature. It has been refined several times in order to get better fit to the result produced through straightforward numerical solution of Boltzmann equations.\\ 

The present study focuses on the solution of Boltzmann equations for leptogenesis in a non-standard (fast expansion) cosmological scenario. The modified set of Boltzmann equations have been solved analytically and the consistency of these analytical expressions has been
checked through comparison with actual numerical solution.
Eventually it has been noticed that the lower bound (eq.\ref{bound_M11}) on lightest RH neutrino mass is reduced further due to the modification of cosmology. This phenomenon has been studied with greater emphasis to find out the modified `lower bound on lightest RH neutrino mass' in case of faster expanding Universe.\\

It is easy to apprehend that the Boltzmann equations governing the evolution of asymmetry will experience non-trivial modification when the expansion rate of the Universe gets altered (from the standard one) due to presence of a new scalar field in the thermal bath. Recently there have been few nice attempts to solve these modified Boltzmann equations numerically. The numerical method is quite efficient in delivering accurate result. However it is not so competent in handling certain problems where scan of a multidimensional parameter space is required, since the time consumed in finishing the task is huge. The time required for this repetitive task of solving the Boltzmann equations for a large number of sample points can be reduced drastically if we have analytical solution for the Boltzmann equations which can express the final asymmetry in terms of some explicit functions of the model parameters. Our present work is primarily focused on the development of the analytical solution of modified Boltzmann equations. We have successfully derived the said approximate analytical expressions and checked their accuracy critically through comparison with actual numerical solution.\\

Modification of the expansion rate of the Universe directly affects the decay parameter $K$ by effectively reducing its value. So the washout effect also decreases. In type-I seesaw leptogenesis, the production of asymmetry (precisely the magnitude of CP asymmetry) mainly depends upon the mass of the lightest RH neutrino apart from the relative phases of the complex Yukawa matrices. In case of vanishing initial abundance, as long as effective value of the decay parameter remains big enough to promptly drive $N_1$ abundance to its equilibrium value, production of asymmetry will not be influenced by cosmology, i.e for a given neutrino mass model same amount of asymmetry will be produced by the decay of $N_1$ in standard as well as modified cosmology. In contrary the evolution of this asymmetry down to present day temperature will be different when the expansion rate of the Universe is faster than the standard case. The produced asymmetry undergoes weaker washout in faster expansion case and as a consequence greater amount of asymmetry survives till the present epoch. Alternatively it can be stated that in this case same amount of final baryon asymmetry can be produced by lighter $N_1$ compared to that of standard cosmology, i.e the existing lower bound (`Davidson-Ibarra bound') on lightest RH neutrino mass (eq.\ref{bound_M11}) can be relaxed without resorting to flavour effects or resonant leptogenesis. Authenticity of this statement has been corroborated in our study through graphical representation of appropriate parameters, where it is vividly demonstrated that for larger portion of the parameter space the said lower bound indeed decreases. The relevant plots (Fig.\ref{delta1_delta2}, Fig.\ref{contour_Delta}) clearly indicate that, for our chosen set of parameters the maximum achievable relaxation on lower bound on $N_1$ mass is approximately $1/30$ times compared to that obtained using standard cosmology. 
\newpage
\appendix
\section{Integration of the inverse decay term in modified cosmology } \label{wid_int}
The integration of the inverse decay term (on the exponent of the efficiency factor) is given by
\begin{eqnarray}
\int \limits^{z }_{z_{in}} \overline{W}^{\prime}_{\rm ID}(z^{\prime\prime}) dz^{\prime\prime} =
\int \limits^{z }_{z_{in}} dz^{\prime\prime} \frac{\overline{z}}{z^{\prime\prime}} W_{\rm ID}(z^{\prime\prime}) f(z^{\prime\prime})~.
\end{eqnarray}
Now we treat $f(z^{\prime\prime})$ to be a constant throughout the interval. It will not affect the final result very much since the maximum contribution to the integral comes from a narrow region around $z=z_B$. In this narrow region the quantity $f(z^{\prime\prime})$ remains nearly constant with a value $f(z_B)$. Thus taking the nearly constant quantity $f(z^{\prime\prime})(\simeq f(z_B))$  outside, the integration can be simplified as
\begin{eqnarray}
 \int \limits^{z }_{z_{in}} \overline{W}^{\prime}_{\rm ID}(z^{\prime\prime}) dz^{\prime\prime} & = &
 f(z_B)\int \limits^{z }_{z_{in}} dz^{\prime\prime} \frac{\overline{z}}{z^{\prime\prime}} W_{\rm ID}(z^{\prime\prime}) \nonumber\\
  &=& f(z_B)\int \limits^{z }_{z_{in}} dz^{\prime\prime} \frac{\overline{z}}{z^{\prime\prime}} 
 \left( \frac{1}{4}K {z^{\prime\prime}}^3 \mathcal{K}_1(z^{\prime\prime}) \right) \nonumber \\
  &=&-f(z_B) \frac{K \overline{z}}{4} \int \limits^{z }_{z_{in}} \frac{d}{d z^{\prime\prime}} \left({z^{\prime\prime}}^2 \mathcal{K}_2(z^{\prime\prime}) \right) dz^{\prime\prime}\nonumber\\
  &=& -f(z_B) \frac{K \overline{z}}{4} \left( z^2\mathcal{K}_2(z) - z^2_{in}\mathcal{K}_2(z_{in}) \right)\nonumber\\
  &=& -f(z_B) \frac{K \overline{z}}{4} \left( z^2\mathcal{K}_2(z) - 2 \right)~.  
\end{eqnarray}
We have used the approximation $z^2_{in}\mathcal{K}_2(z_{in})\rightarrow 2$ (since $z_{in}$ is very small) for the simplification of last but one step.
When we compute the final efficiency factor $\kappa_f$, the above integration has to be evaluated up to a very large value of $z$ (theoretically $z \rightarrow \infty$, for which $z^2\mathcal{K}_2(z) \rightarrow 0$). In that case the integral under consideration reduces to
\begin{equation}
  \int \limits^{z \rightarrow \infty }_{z_{in} \rightarrow 0} \overline{W}^{\prime}_{\rm ID}(z^{\prime\prime}) dz^{\prime\prime}
  = f(z_B) \frac{K \overline{z_B}(K)}{2}~.
\end{equation}
\section{Evolution of RH neutrino abundance ($\eta_N(z)$) in modified cosmology } \label{eta_N_mod}
We start from the first Boltzmann equation which governs the evolution of RH neutrino abundance, i.e
\begin{equation}
 \frac{d \eta_{N}}{dz}=- D(z)  \bigg (\eta_{N}(z)-\eta^{\rm eq}_{N}(z)  \bigg) 
\end{equation}
where 
\begin{eqnarray}
 D(z)&=&z K_{\rm eff}(z) \frac{\mathcal{K}_1(z)}{\mathcal{K}_2(z)} \nonumber\\
     &=& z K \left[ 1+ \left(\frac{\gamma}{z} \right)^{\delta} \right]^{-1/2} \frac{\mathcal{K}_1(z)}{\mathcal{K}_2(z)}~.
\end{eqnarray}
Since we are concerned about the behaviour of $\eta_N(z)$ at some early epoch (before it has reached equilibrium), $z$ in our case is very
small, i.e $z<z_{eq}(<1)$. It allows us to take $\eta^{\rm eq}_{N}(z)=3/4$. Using this constant value of equilibrium RH neutrino abundance along with modified expression of $D(z)$, the above mentioned Boltzmann equation can be written in the form
\begin{eqnarray}
\frac{d \eta_{N}(z)}{\eta_{N}(z)-3/4}= z K \left[ 1+ \left(\frac{\gamma}{z} \right)^{\delta} \right]^{-1/2} \frac{\mathcal{K}_1(z)}{\mathcal{K}_2(z)} d z ~.\label{diff_eqn}
\end{eqnarray}
In the present case $z \ll 1$, which permits us to use the approximate expression for the ratio\cite{Buchmuller:2004nz} of the Bessel functions as
\begin{eqnarray}
 \frac{\mathcal{K}_2(z)}{\mathcal{K}_1(z)} =  \frac{1}{z} \left( \frac{15}{8} + z \right) \simeq \frac{2}{z}~.
\end{eqnarray}
The ratio $\gamma/z \gg 1$, which implies that $1+(\gamma/z)^\delta \simeq (\gamma/z)^\delta$. The differential equation (\ref{diff_eqn}) reduces to a simpler form after incorporation of the said approximations, i.e
\begin{equation}
 \frac{d \eta_{N}(z)}{\eta_{N}(z)-3/4} \simeq - K \frac{z^2}{2} \gamma^{-\delta/2} z^{\delta/2} d z = 
 -\left( \frac{K \gamma^{-\delta/2}}{2} \right) z^{2+\delta/2} dz ~.
\end{equation}
A straightforward integration gives
\begin{equation}
 \ln \left[ \eta_{N}(z)-3/4 \right] = - \frac{K \gamma^{-\delta/2}}{6+\delta} z^{(6+\delta)/2} + C~.
\end{equation}
The boundary condition $\eta_{N}(z=0)=0$, enables us to find the integration constant $C=\ln \left( -3/4 \right)$. Now we are at a stage to write the analytical expression of $\eta_{N}(z)$ using the value of the integration constant in the above equation, followed by exponentiation of the R.H.S., i.e 
\begin{eqnarray}
 \frac{\eta_{N}(z)-3/4}{-3/4} &=& \exp \left( - \frac{K \gamma^{-\delta/2}}{6+\delta} z^{(6+\delta)/2} \right)~, \nonumber\\
 \Rightarrow \eta_{N}(z) &=& \frac{3}{4} \left[1- \exp \left( - \frac{K \gamma^{-\delta/2}}{6+\delta} z^{(6+\delta)/2} \right) \right]~.
\end{eqnarray}
\section{Expression for $z_{eq}$ in case of faster expansion ($\delta>2$)}\label{zeq_mod}
The first Boltzmann equation can be solved systematically\cite{Buchmuller:2004nz} in powers of $K$ to express the difference 
$\tilde{\Delta}(z)(=\eta_{N}(z)-\eta^{\rm eq}_{N}(z)  )$ as
\begin{eqnarray}
 \tilde{\Delta}(z) & = & -\frac{1}{D(z)} \frac{d \eta^{\rm eq}_{N}(z)}{dz} + \mathcal{O}\left(\frac{1}{K^2}\right) \\
                   & \simeq & -\frac{\mathcal{K}_2(z)}{K z f(z) \mathcal{K}_1(z)} \left( -\frac{3}{8} z^2 \mathcal{K}_1(z) \right) \nonumber\\
                   & = & \frac{3}{8K} \frac{z \mathcal{K}_2(z)}{f(z)} ~.\label{diff}
\end{eqnarray}
Although $z_{eq} \gg 1$, it is much less than the value of $\gamma$ and thus the approximation of $f(z) \simeq (\gamma/z)^{-\delta/2}$ holds good. Using this approximate expression of $f(z)$ in eq.\ref{diff}, the ratio of difference $\tilde{\Delta}(z)$ to equilibrium $N_1$ density comes out be
\begin{equation}
R_{eq}(z)=\frac{\tilde{\Delta}(z)}{\eta^{\rm eq}_{N}(z)} \simeq \frac{\frac{3}{8K}\left(\frac{\gamma}{z}\right)^{\delta/2}z \mathcal{K}_2(z)}
{\frac{3}{8}z^2  \mathcal{K}_2(z)} = \frac{\gamma^{\delta/2}}{K z^{1+\delta/2}} ~.
\end{equation}
Ideally equilibrium is marked by a point $(z)$ for which $N_1$ abundance touches its equilibrium value, i.e $\tilde{\Delta}(z)=0$. In practice 
we may demand equilibrium has been reached when the above mentioned ratio becomes smaller than a certain limit. Mathematical representation of this statement is $R_{eq}(z) \leq 1/p$, where $p$ is a big number (we have checked that it is sufficient to take $p \sim 10$). Therefore the point $z_{eq}$ can be found by solving the equation
\begin{eqnarray}
 && \frac{\tilde{\Delta}(z)}{\eta^{\rm eq}_{N}(z)}  =  \frac{1}{p} ~,\\
 && \Rightarrow \frac{K z^{1+\delta/2}}{\gamma^{\delta/2}}  =  p ~,\nonumber \\
 && \Rightarrow z  =  \left( \frac{p \gamma^{\delta/2} }{K} \right)^{\frac{2}{2+\delta}} = z_{eq}~.
\end{eqnarray}
\section{Integration of the inverse decay term in modified cosmology ($\delta > 2$) } \label{wid_int1}
Before entering into the evaluation of actual integral let us recall the definition of upper incomplete gamma function (eq.\ref{incomp_gamma}), from which the difference of the gamma functions evaluated at two points $x_1,x_2$ (with $x_1<x_2$) is computed as 
\begin{eqnarray}
 \Gamma(a,x_1) -\Gamma(a,x_2) & = & \int \limits^{ \infty }_{ x_1 } e^{-t}~ t^{a-1} dt ~ - ~\int \limits^{ \infty }_{ x_2 } e^{-t}~ t^{a-1} dt \nonumber\\
 & = & \int \limits^{ x^\prime }_{ x_1 } I(t)dt + \int \limits^{ \infty }_{ x^\prime } I(t)dt - \int \limits^{ x^\prime }_{ x_2 } I(t)dt
 -\int \limits^{ \infty }_{ x^\prime } I(t)dt ~~~({\rm using}~I(t)=e^{-t}~ t^{a-1})\nonumber \\
 & = & \int \limits^{ x^\prime }_{ x_1 } I(t)dt + \int \limits^{ x_2 }_{ x^\prime } I(t)dt \nonumber\\
 & = & \int \limits^{ x_2 }_{ x_1 } I(t)dt ~.
\end{eqnarray}
Since we are dealing with larger values of $\delta$ in this section, the $z$ values of our interest is also much greater than unity. It allows us to use the approximation
\begin{equation}
 \mathcal{K}_1(z)  = \frac{1}{z} \sqrt{1+\frac{\pi}{2}z} e^{-z} \simeq \frac{1}{z} \sqrt{\frac{\pi}{2}z} e^{-z} ~.\label{k1_approx}
\end{equation}
In spite of bigger values of $z$ the approximate form of $f(z)$ is
\begin{equation}
f(z) = \left[ 1+ \left(\frac{\gamma}{z} \right)^\delta \right]^{-1/2} \simeq  \left(\frac{\gamma}{z} \right)^{-\delta/2}~. \label{f_approx}
\end{equation}
Let us now evaluate the integral, i.e 
\begin{eqnarray}
\int \limits^{z }_{z^\prime} W^{\prime}_{\rm ID}(z^{\prime\prime}) dz^{\prime\prime} 
& = & \int \limits^{z }_{z^\prime} W_{\rm ID}(z^{\prime\prime}) f(z^{\prime\prime}) dz^{\prime\prime} \nonumber \\
& = & \int \limits^{z }_{z^\prime} \frac{1}{4} K {z^{\prime\prime}}^3 \mathcal{K}_1(z^{\prime\prime}) f(z^{\prime\prime}) dz^{\prime\prime}  \nonumber \\
& = & \int \limits^{z }_{z^\prime} \frac{1}{4} K {z^{\prime\prime}}^3 \frac{1}{z^{\prime\prime}} \sqrt{\frac{\pi}{2}z^{\prime\prime}} e^{-{z}^{\prime\prime}} \left(\frac{\gamma}{z^{\prime\prime}} \right)^{-\delta/2} ~~({\rm using ~ eq.\ref{k1_approx},~eq.\ref{f_approx}}) \nonumber \\
& = & \frac{K}{4}\sqrt{\frac{\pi}{2}} \gamma^{-\delta/2} \int \limits^{z }_{z^\prime} {z^{\prime\prime}}^{\frac{5+\delta}{2}} e^{-{z}^{\prime\prime}} d z^{\prime\prime} \nonumber \\
& = & C \int \limits^{z }_{z^\prime} \left({z^{\prime\prime}}\right)^{a-1} e^{-{z}^{\prime\prime}} d z^{\prime\prime} 
~~~({\rm where }~C=\frac{K}{4}\sqrt{\frac{\pi}{2}} \gamma^{-\delta/2},~a=\frac{7+\delta}{2})\nonumber \\
& = & C \left[ \Gamma(a,z^\prime) - \Gamma(a,z) \right]~.
\end{eqnarray}
\section{Effect of modified cosmology in weak washout regime}\label{weak_wash}
We divide this discussion in two parts depending upon the initial condition, i.e the initial RH neutrino abundance.\\

Let us first explore the `thermal initial abundance case ($\eta_{N}(z=z_{in})=3/4$)'. 
The washout term (mainly due to inverse decay) is given by
\begin{eqnarray}
 W^{\prime}_{\rm ID}(z) &=&\frac{1}{4} K_{\rm eff} z^3 \mathcal{K}_1(z) \nonumber\\
                        &=& \frac{1}{4} \frac{K}{\sqrt{1+\left(\frac{\gamma}{z} \right)^\delta}} z^3 \mathcal{K}_1(z)~.
\end{eqnarray}
Since we are working in the weak washout regime, $K$ is already very small. The presence of the factor in the denominator makes it smaller. So the washout term $( W^{\prime}_{\rm ID})$ as a whole becomes smaller than that of the standard case. Therefore the efficiency factor is simplified to
\begin{eqnarray}
 \kappa(z) &=& -\frac{4}{3}\int \limits^{z }_{z_{in}} d z^{\prime} \frac{d {\eta_N}(z^\prime)}{d z^{\prime}} ~
e^{- \int \limits^{z }_{z^\prime} W^\prime_{\rm ID}(z^{\prime\prime}) dz^{\prime\prime}} ~\nonumber\\
           & \simeq & -\frac{4}{3}\int \limits^{z }_{z_{in}} d z^{\prime} \frac{d {\eta_N}(z^\prime)}{d z^{\prime}}~~~~~~({\rm neglecting ~the} ~W^\prime_{\rm ID} ~{\rm term}) \nonumber\\
           & = &  -\frac{4}{3} \left[{\eta_N}(z) -{\eta_N}(z_{in})  \right]~,
\end{eqnarray}
from which the final efficiency factor can be obtained as
\begin{eqnarray}
\kappa_f &=&\kappa(z\rightarrow\infty) \nonumber\\
         &=& -\frac{4}{3} \left[{\eta_N}(z\rightarrow\infty) -{\eta_N}(z_{in}) \right] \nonumber\\
         &=& -\frac{4}{3} \left[0 -\frac{3}{4} \right] =1~.
\end{eqnarray}
The efficiency factor remains the same as it was in the standard radiation dominated case.\\

We now move to the case of \textquoteleft ~dynamical initial abundance \textquoteright. Here the initial RH neutrino abundance is vanishing and it is generated dynamically due to the inverse decay of RH neutrinos. The total efficiency factor should be the sum of positive and negative contribution, i.e
\begin{eqnarray}
\kappa(z) &=& \kappa^+(z) + \kappa^-(z) \nonumber\\
          &=& -\frac{4}{3}\int \limits^{z }_{z_{eq}} d z^{\prime} \frac{d {\eta_N}(z^\prime)}{d z^{\prime}} ~
e^{- \int \limits^{z }_{z^\prime} W^\prime_{\rm ID}(z^{\prime\prime}) dz^{\prime\prime}} 
-\frac{4}{3}\int \limits^{z_{eq} }_{z_{in}} d z^{\prime} \frac{d {\eta_N}(z^\prime)}{d z^{\prime}} ~
e^{- \int \limits^{z }_{z^\prime} W^\prime_{\rm ID}(z^{\prime\prime}) dz^{\prime\prime}} \nonumber\\
          &=& \frac{4}{3} \left[{\eta_N}(z_{eq}) -{\eta_N}(z) \right] -2 \left[1-\exp\left(-\frac{2}{3}{\eta_N}(z_{eq}) \right) \right]  ~.
\end{eqnarray}
In standard cosmology we get ${\eta_N}(K)={\eta_N}(z_{eq})=(9\pi K)/16$. Therefore the final efficiency factor in standard cosmology is found to be
\begin{eqnarray}
\kappa_f &=&\kappa (z \rightarrow\infty) \nonumber\\
         &=& \frac{4}{3} {\eta_N}(K) - 2\left(1-1 + \frac{2}{3}{\eta_N}(K)-\frac{1}{2}\frac{4}{9}{\eta_N}^2(K) +...\right) \nonumber\\
         & \simeq & \frac{4}{9}{\eta_N}^2(K) ~~~~({\rm neglecting~higher}~(>2)~{\rm order~terms~in}~K)\label{kf_weak}\\
         &=& \frac{9 \pi^2 K^2}{64}~,
\end{eqnarray}
We now try to calculate ${\eta_N}(z_{eq})$ in case of modified cosmology. Let us start from the simplified form of first Boltzmann equation
\begin{eqnarray}
 \frac{d \eta_{N}(z)}{dz} &\simeq& D(z) \eta^{eq}_{N}(z) ~~~~~~~ ({\rm neglecting} ~\eta_{N}(z)~{\rm due ~ to ~initial~condition}) \nonumber \\
 & = & z K \left[ 1+\left(\frac{\gamma}{z} \right)^\delta \right]^{-1/2} \frac{\mathcal{K}_1(z)}{\mathcal{K}_2(z)}
       ~\frac{3}{8} z^2 \mathcal{K}_2(z) ~.
\end{eqnarray}
Since we are dealing with weak washout, equilibrium will be reached for a high value of $z$ ($z \gg 1$). Using the approximate forms of $\mathcal{K}_1(z)$ and $f(z)$ (eq.\ref{k1_approx}, eq.\ref{f_approx}) in above differential equation we get 
\begin{eqnarray}
 \frac{d \eta_{N}(z)}{dz} &\simeq & \frac{3}{8} K \left(\frac{\gamma}{z} \right)^{-\delta/2} z^2 \sqrt{\frac{\pi}{2}}z^{1/2} e^{-z} \nonumber \\
& = & \frac{3 K}{8} \sqrt{\frac{\pi}{2}} \gamma^{-\delta/2} z^{\frac{5+\delta}{2}} e^{-z} ~.
\end{eqnarray}
Integrating above differential equation from very small value of $z(\sim 0)$ to $z_{eq}(\gg 1)$
\begin{eqnarray}
 \int \limits^{z_{eq} \rightarrow \infty }_{z_{in} \rightarrow 0} d {\eta_N}(z) 
 &=&  \frac{3 K}{8} \sqrt{\frac{\pi}{2}} \gamma^{-\delta/2} \int \limits^{z_{eq} \rightarrow \infty }_{z_{in} \rightarrow 0}  z^{\frac{7+\delta}{2}-1} e^{-z} dz \nonumber \\
 \Rightarrow {\eta_N}(z_{eq}) &=& \frac{3 K}{8} \sqrt{\frac{\pi}{2}} \gamma^{-\delta/2} \Gamma \left( \frac{7+\delta}{2} \right) =  {\eta^\prime_N}(K)~. 
\end{eqnarray}
It follows from eq.\ref{kf_weak} that the final efficiency factor in this case is given by
\begin{equation}
\kappa^\prime_f =  \frac{4}{9}{\eta^\prime_N}^2(K)~.
\end{equation}
To examine the change in efficiency factor due to introduction of faster expansion we compute the ratio
\begin{eqnarray}
\frac{\kappa^\prime_f}{\kappa_f} &=& \frac{\left({\eta^\prime_N}(K)\right)^2}{\left({\eta_N}(K)\right)^2} \nonumber\\
                                 &=& \left( \frac{\gamma^{-\delta/2} \Gamma \left( \frac{7+\delta}{2} \right)}{\Gamma \left( \frac{7}{2} \right)} \right)^2 ~.
\end{eqnarray}
Although $\Gamma \left( \frac{7+\delta}{2} \right) > \Gamma \left( \frac{7}{2} \right)$, presence of the term $\gamma^{-\delta/2}(\ll 1 ~{\rm with~}\gamma=50,\delta \gtrsim 1)$ in the numerator makes the overall ratio $\ll 1$, which implies $\kappa^\prime_f < \kappa_f$. The value of final efficiency factor becomes smaller than the standard radiation dominated scenario. Thus it will become more difficult to produce the observed asymmetry. Thus it is justified to comment that modified cosmology (or faster expansion of Universe) does not have any noticeable effect
in the process of baryogenesis through leptogenesis in weak washout regime.

\acknowledgments
M.C would like to acknowledge the financial support provided by SERB-DST, Govt. of India through the NPDF project {\bf PDF/2019/000622}. We thank Sougata Ganguly and Soumitra SenGupta for many helpful discussions.

\bibliographystyle{JHEP}
\bibliography{baryasy_modcosmo_v2.bib}

\providecommand{\href}[2]{#2}\begingroup\raggedright\begin{thebibliography}{10}

\bibitem{Pontecorvo:1967fh}
B.~Pontecorvo, \emph{{Neutrino Experiments and the Problem of Conservation of
  Leptonic Charge}}, {\emph{Zh. Eksp. Teor. Fiz.} {\bfseries 53} (1967) 1717}.

\bibitem{Gribov:1968kq}
V.~N. Gribov and B.~Pontecorvo, \emph{{Neutrino astronomy and lepton charge}},
  \href{https://doi.org/10.1016/0370-2693(69)90525-5}{\emph{Phys. Lett. B}
  {\bfseries 28} (1969) 493}.

\bibitem{Esteban:2018azc}
I.~Esteban, M.~C. Gonzalez-Garcia, A.~Hernandez-Cabezudo, M.~Maltoni and
  T.~Schwetz, \emph{{Global analysis of three-flavour neutrino oscillations:
  synergies and tensions in the determination of $\theta_{23}$, $\delta_{CP}$,
  and the mass ordering}},
  \href{https://doi.org/10.1007/JHEP01(2019)106}{\emph{JHEP} {\bfseries 01}
  (2019) 106} [\href{https://arxiv.org/abs/1811.05487}{{\ttfamily
  1811.05487}}].

\bibitem{Abe:2017vif}
{\scshape T2K} collaboration, \emph{{Measurement of neutrino and antineutrino
  oscillations by the T2K experiment including a new additional sample of
  $\nu_e$ interactions at the far detector}},
  \href{https://doi.org/10.1103/PhysRevD.96.092006}{\emph{Phys. Rev. D}
  {\bfseries 96} (2017) 092006}
  [\href{https://arxiv.org/abs/1707.01048}{{\ttfamily 1707.01048}}].

\bibitem{Abe:2018wpn}
{\scshape T2K} collaboration, \emph{{Search for CP Violation in Neutrino and
  Antineutrino Oscillations by the T2K Experiment with $2.2\times10^{21}$
  Protons on Target}},
  \href{https://doi.org/10.1103/PhysRevLett.121.171802}{\emph{Phys. Rev. Lett.}
  {\bfseries 121} (2018) 171802}
  [\href{https://arxiv.org/abs/1807.07891}{{\ttfamily 1807.07891}}].

\bibitem{Adamson:2017gxd}
{\scshape NOvA} collaboration, \emph{{Constraints on Oscillation Parameters
  from $\nu_e$ Appearance and $\nu_\mu$ Disappearance in NOvA}},
  \href{https://doi.org/10.1103/PhysRevLett.118.231801}{\emph{Phys. Rev. Lett.}
  {\bfseries 118} (2017) 231801}
  [\href{https://arxiv.org/abs/1703.03328}{{\ttfamily 1703.03328}}].

\bibitem{NOvA:2018gge}
{\scshape NOvA} collaboration, \emph{{New constraints on oscillation parameters
  from $\nu_e$ appearance and $\nu_\mu$ disappearance in the NOvA experiment}},
  \href{https://doi.org/10.1103/PhysRevD.98.032012}{\emph{Phys. Rev. D}
  {\bfseries 98} (2018) 032012}
  [\href{https://arxiv.org/abs/1806.00096}{{\ttfamily 1806.00096}}].

\bibitem{Adey:2018zwh}
{\scshape Daya Bay} collaboration, \emph{{Measurement of the Electron
  Antineutrino Oscillation with 1958 Days of Operation at Daya Bay}},
  \href{https://doi.org/10.1103/PhysRevLett.121.241805}{\emph{Phys. Rev. Lett.}
  {\bfseries 121} (2018) 241805}
  [\href{https://arxiv.org/abs/1809.02261}{{\ttfamily 1809.02261}}].

\bibitem{Bak:2018ydk}
{\scshape RENO} collaboration, \emph{{Measurement of Reactor Antineutrino
  Oscillation Amplitude and Frequency at RENO}},
  \href{https://doi.org/10.1103/PhysRevLett.121.201801}{\emph{Phys. Rev. Lett.}
  {\bfseries 121} (2018) 201801}
  [\href{https://arxiv.org/abs/1806.00248}{{\ttfamily 1806.00248}}].

\bibitem{Adamson:2013whj}
{\scshape MINOS} collaboration, \emph{{Measurement of Neutrino and Antineutrino
  Oscillations Using Beam and Atmospheric Data in MINOS}},
  \href{https://doi.org/10.1103/PhysRevLett.110.251801}{\emph{Phys. Rev. Lett.}
  {\bfseries 110} (2013) 251801}
  [\href{https://arxiv.org/abs/1304.6335}{{\ttfamily 1304.6335}}].

\bibitem{Abe:2014bwa}
{\scshape Double Chooz} collaboration, \emph{{Improved measurements of the
  neutrino mixing angle $\theta_{13}$ with the Double Chooz detector}},
  \href{https://doi.org/10.1007/JHEP02(2015)074}{\emph{JHEP} {\bfseries 10}
  (2014) 086} [\href{https://arxiv.org/abs/1406.7763}{{\ttfamily 1406.7763}}].

\bibitem{Anton:2019wmi}
{\scshape EXO-200} collaboration, \emph{{Search for Neutrinoless Double-$\beta$
  Decay with the Complete EXO-200 Dataset}},
  \href{https://doi.org/10.1103/PhysRevLett.123.161802}{\emph{Phys. Rev. Lett.}
  {\bfseries 123} (2019) 161802}
  [\href{https://arxiv.org/abs/1906.02723}{{\ttfamily 1906.02723}}].

\bibitem{Agostini:2018tnm}
{\scshape GERDA} collaboration, \emph{{Improved Limit on Neutrinoless
  Double-$\beta$ Decay of $^{76}$Ge from GERDA Phase II}},
  \href{https://doi.org/10.1103/PhysRevLett.120.132503}{\emph{Phys. Rev. Lett.}
  {\bfseries 120} (2018) 132503}
  [\href{https://arxiv.org/abs/1803.11100}{{\ttfamily 1803.11100}}].

\bibitem{Gando:2018kyv}
{\scshape Kamland-Zen} collaboration, \emph{{Neutrinoless double beta decay
  search with KamLAND-Zen}},
  \href{https://doi.org/10.22323/1.337.0068}{\emph{PoS} {\bfseries NOW2018}
  (2018) 068}.

\bibitem{Alduino:2017ehq}
{\scshape CUORE} collaboration, \emph{{First Results from CUORE: A Search for
  Lepton Number Violation via $0\nu\beta\beta$ Decay of $^{130}$Te}},
  \href{https://doi.org/10.1103/PhysRevLett.120.132501}{\emph{Phys. Rev. Lett.}
  {\bfseries 120} (2018) 132501}
  [\href{https://arxiv.org/abs/1710.07988}{{\ttfamily 1710.07988}}].

\bibitem{Azzolini:2019tta}
{\scshape CUPID} collaboration, \emph{{Final result of CUPID-0 phase-I in the
  search for the $^{82}$Se Neutrinoless Double-$\beta$ Decay}},
  \href{https://doi.org/10.1103/PhysRevLett.123.032501}{\emph{Phys. Rev. Lett.}
  {\bfseries 123} (2019) 032501}
  [\href{https://arxiv.org/abs/1906.05001}{{\ttfamily 1906.05001}}].

\bibitem{WMAP:2003elm}
{\scshape WMAP} collaboration, \emph{{First year Wilkinson Microwave Anisotropy
  Probe (WMAP) observations: Determination of cosmological parameters}},
  \href{https://doi.org/10.1086/377226}{\emph{Astrophys. J. Suppl.} {\bfseries
  148} (2003) 175} [\href{https://arxiv.org/abs/astro-ph/0302209}{{\ttfamily
  astro-ph/0302209}}].

\bibitem{WMAP:2008lyn}
{\scshape WMAP} collaboration, \emph{{Five-Year Wilkinson Microwave Anisotropy
  Probe (WMAP) Observations: Cosmological Interpretation}},
  \href{https://doi.org/10.1088/0067-0049/180/2/330}{\emph{Astrophys. J.
  Suppl.} {\bfseries 180} (2009) 330}
  [\href{https://arxiv.org/abs/0803.0547}{{\ttfamily 0803.0547}}].

\bibitem{WMAP:2008ydk}
{\scshape WMAP} collaboration, \emph{{Five-Year Wilkinson Microwave Anisotropy
  Probe (WMAP) Observations: Data Processing, Sky Maps, and Basic Results}},
  \href{https://doi.org/10.1088/0067-0049/180/2/225}{\emph{Astrophys. J.
  Suppl.} {\bfseries 180} (2009) 225}
  [\href{https://arxiv.org/abs/0803.0732}{{\ttfamily 0803.0732}}].

\bibitem{WMAP:2010qai}
{\scshape WMAP} collaboration, \emph{{Seven-Year Wilkinson Microwave Anisotropy
  Probe (WMAP) Observations: Cosmological Interpretation}},
  \href{https://doi.org/10.1088/0067-0049/192/2/18}{\emph{Astrophys. J. Suppl.}
  {\bfseries 192} (2011) 18} [\href{https://arxiv.org/abs/1001.4538}{{\ttfamily
  1001.4538}}].

\bibitem{WMAP:2012nax}
{\scshape WMAP} collaboration, \emph{{Nine-Year Wilkinson Microwave Anisotropy
  Probe (WMAP) Observations: Cosmological Parameter Results}},
  \href{https://doi.org/10.1088/0067-0049/208/2/19}{\emph{Astrophys. J. Suppl.}
  {\bfseries 208} (2013) 19} [\href{https://arxiv.org/abs/1212.5226}{{\ttfamily
  1212.5226}}].

\bibitem{Planck:2018vyg}
{\scshape Planck} collaboration, \emph{{Planck 2018 results. VI. Cosmological
  parameters}},
  \href{https://doi.org/10.1051/0004-6361/201833910}{\emph{Astron. Astrophys.}
  {\bfseries 641} (2020) A6}
  [\href{https://arxiv.org/abs/1807.06209}{{\ttfamily 1807.06209}}].

\bibitem{Planck:2013pxb}
{\scshape Planck} collaboration, \emph{{Planck 2013 results. XVI. Cosmological
  parameters}},
  \href{https://doi.org/10.1051/0004-6361/201321591}{\emph{Astron. Astrophys.}
  {\bfseries 571} (2014) A16}
  [\href{https://arxiv.org/abs/1303.5076}{{\ttfamily 1303.5076}}].

\bibitem{Aghanim:2016yuo}
{\scshape Planck} collaboration, \emph{{Planck intermediate results. XLVI.
  Reduction of large-scale systematic effects in HFI polarization maps and
  estimation of the reionization optical depth}},
  \href{https://doi.org/10.1051/0004-6361/201628890}{\emph{Astron. Astrophys.}
  {\bfseries 596} (2016) A107}
  [\href{https://arxiv.org/abs/1605.02985}{{\ttfamily 1605.02985}}].

\bibitem{Riotto:1998bt}
A.~Riotto, \emph{{Theories of baryogenesis}},  in \emph{{ICTP Summer School in
  High-Energy Physics and Cosmology}}, pp.~326--436, 7, 1998,
  \href{https://arxiv.org/abs/hep-ph/9807454}{{\ttfamily hep-ph/9807454}}.

\bibitem{Cline:2006ts}
J.~M. Cline, \emph{{Baryogenesis}},  in \emph{{Les Houches Summer School -
  Session 86: Particle Physics and Cosmology: The Fabric of Spacetime}}, 9,
  2006, \href{https://arxiv.org/abs/hep-ph/0609145}{{\ttfamily
  hep-ph/0609145}}.

\bibitem{Dine:2003ax}
M.~Dine and A.~Kusenko, \emph{{The Origin of the matter - antimatter
  asymmetry}}, \href{https://doi.org/10.1103/RevModPhys.76.1}{\emph{Rev. Mod.
  Phys.} {\bfseries 76} (2003) 1}
  [\href{https://arxiv.org/abs/hep-ph/0303065}{{\ttfamily hep-ph/0303065}}].

\bibitem{Fukugita:1986hr}
M.~Fukugita and T.~Yanagida, \emph{{Baryogenesis Without Grand Unification}},
  \href{https://doi.org/10.1016/0370-2693(86)91126-3}{\emph{Phys. Lett. B}
  {\bfseries 174} (1986) 45}.

\bibitem{Riotto:1999yt}
A.~Riotto and M.~Trodden, \emph{{Recent progress in baryogenesis}},
  \href{https://doi.org/10.1146/annurev.nucl.49.1.35}{\emph{Ann. Rev. Nucl.
  Part. Sci.} {\bfseries 49} (1999) 35}
  [\href{https://arxiv.org/abs/hep-ph/9901362}{{\ttfamily hep-ph/9901362}}].

\bibitem{Buchmuller:2004nz}
W.~Buchmuller, P.~Di~Bari and M.~Plumacher, \emph{{Leptogenesis for
  pedestrians}}, \href{https://doi.org/10.1016/j.aop.2004.02.003}{\emph{Annals
  Phys.} {\bfseries 315} (2005) 305}
  [\href{https://arxiv.org/abs/hep-ph/0401240}{{\ttfamily hep-ph/0401240}}].

\bibitem{Davidson:2008bu}
S.~Davidson, E.~Nardi and Y.~Nir, \emph{{Leptogenesis}},
  \href{https://doi.org/10.1016/j.physrep.2008.06.002}{\emph{Phys. Rept.}
  {\bfseries 466} (2008) 105}
  [\href{https://arxiv.org/abs/0802.2962}{{\ttfamily 0802.2962}}].

\bibitem{Bertuzzo:2009im}
E.~Bertuzzo, P.~Di~Bari, F.~Feruglio and E.~Nardi, \emph{{Flavor symmetries,
  leptogenesis and the absolute neutrino mass scale}},
  \href{https://doi.org/10.1088/1126-6708/2009/11/036}{\emph{JHEP} {\bfseries
  11} (2009) 036} [\href{https://arxiv.org/abs/0908.0161}{{\ttfamily
  0908.0161}}].

\bibitem{DiBari:2012fz}
P.~Di~Bari, \emph{{An introduction to leptogenesis and neutrino properties}},
  \href{https://doi.org/10.1080/00107514.2012.701096}{\emph{Contemp. Phys.}
  {\bfseries 53} (2012) 315} [\href{https://arxiv.org/abs/1206.3168}{{\ttfamily
  1206.3168}}].

\bibitem{Adhikary:2014qba}
B.~Adhikary, M.~Chakraborty and A.~Ghosal, \emph{{Flavored leptogenesis with
  quasidegenerate neutrinos in a broken cyclic symmetric model}},
  \href{https://doi.org/10.1103/PhysRevD.93.113001}{\emph{Phys. Rev. D}
  {\bfseries 93} (2016) 113001}
  [\href{https://arxiv.org/abs/1407.6173}{{\ttfamily 1407.6173}}].

\bibitem{Samanta:2016hcj}
R.~Samanta, M.~Chakraborty, P.~Roy and A.~Ghosal, \emph{{Baryon asymmetry via
  leptogenesis in a neutrino mass model with complex scaling}},
  \href{https://doi.org/10.1088/1475-7516/2017/03/025}{\emph{JCAP} {\bfseries
  03} (2017) 025} [\href{https://arxiv.org/abs/1610.10081}{{\ttfamily
  1610.10081}}].

\bibitem{Samanta:2018hqm}
R.~Samanta and M.~Chakraborty, \emph{{A study on a minimally broken residual
  TBM-Klein symmetry with its implications on flavoured leptogenesis and ultra
  high energy neutrino flux ratios}},
  \href{https://doi.org/10.1088/1475-7516/2019/02/003}{\emph{JCAP} {\bfseries
  02} (2019) 003} [\href{https://arxiv.org/abs/1802.04751}{{\ttfamily
  1802.04751}}].

\bibitem{Samanta:2018efa}
R.~Samanta, R.~Sinha and A.~Ghosal, \emph{{Importance of generalized $\mu\tau$
  symmetry and its CP extension on neutrino mixing and leptogenesis}},
  \href{https://doi.org/10.1007/JHEP10(2019)057}{\emph{JHEP} {\bfseries 10}
  (2019) 057} [\href{https://arxiv.org/abs/1805.10031}{{\ttfamily
  1805.10031}}].

\bibitem{Affleck:1984fy}
I.~Affleck and M.~Dine, \emph{{A New Mechanism for Baryogenesis}},
  \href{https://doi.org/10.1016/0550-3213(85)90021-5}{\emph{Nucl. Phys. B}
  {\bfseries 249} (1985) 361}.

\bibitem{Dine:1995kz}
M.~Dine, L.~Randall and S.~D. Thomas, \emph{{Baryogenesis from flat directions
  of the supersymmetric standard model}},
  \href{https://doi.org/10.1016/0550-3213(95)00538-2}{\emph{Nucl. Phys. B}
  {\bfseries 458} (1996) 291}
  [\href{https://arxiv.org/abs/hep-ph/9507453}{{\ttfamily hep-ph/9507453}}].

\bibitem{Ignatiev:1978uf}
A.~Y. Ignatiev, N.~V. Krasnikov, V.~A. Kuzmin and A.~N. Tavkhelidze,
  \emph{{Universal CP Noninvariant Superweak Interaction and Baryon Asymmetry
  of the Universe}},
  \href{https://doi.org/10.1016/0370-2693(78)90900-0}{\emph{Phys. Lett. B}
  {\bfseries 76} (1978) 436}.

\bibitem{Ellis:1978xg}
J.~R. Ellis, M.~K. Gaillard and D.~V. Nanopoulos, \emph{{Baryon Number
  Generation in Grand Unified Theories}},
  \href{https://doi.org/10.1016/0370-2693(79)91190-0}{\emph{Phys. Lett. B}
  {\bfseries 80} (1979) 360}.

\bibitem{Minkowski:1977sc}
P.~Minkowski, \emph{{$\mu \to e\gamma$ at a Rate of One Out of $10^{9}$ Muon
  Decays?}}, \href{https://doi.org/10.1016/0370-2693(77)90435-X}{\emph{Phys.
  Lett. B} {\bfseries 67} (1977) 421}.

\bibitem{GellMann:1980vs}
M.~Gell-Mann, P.~Ramond and R.~Slansky, \emph{{Complex Spinors and Unified
  Theories}}, {\emph{Conf. Proc. C} {\bfseries 790927} (1979) 315}
  [\href{https://arxiv.org/abs/1306.4669}{{\ttfamily 1306.4669}}].

\bibitem{Yanagida:1980xy}
T.~Yanagida, \emph{{Horizontal Symmetry and Masses of Neutrinos}},
  \href{https://doi.org/10.1143/PTP.64.1103}{\emph{Prog. Theor. Phys.}
  {\bfseries 64} (1980) 1103}.

\bibitem{Mohapatra:1979ia}
R.~N. Mohapatra and G.~Senjanovic, \emph{{Neutrino Mass and Spontaneous Parity
  Nonconservation}},
  \href{https://doi.org/10.1103/PhysRevLett.44.912}{\emph{Phys. Rev. Lett.}
  {\bfseries 44} (1980) 912}.

\bibitem{Sakharov:1967dj}
A.~D. Sakharov, \emph{{Violation of CP Invariance, C asymmetry, and baryon
  asymmetry of the universe}},
  \href{https://doi.org/10.1070/PU1991v034n05ABEH002497}{\emph{Pisma Zh. Eksp.
  Teor. Fiz.} {\bfseries 5} (1967) 32}.

\bibitem{Davidson:2002qv}
S.~Davidson and A.~Ibarra, \emph{{A Lower bound on the right-handed neutrino
  mass from leptogenesis}},
  \href{https://doi.org/10.1016/S0370-2693(02)01735-5}{\emph{Phys. Lett. B}
  {\bfseries 535} (2002) 25}
  [\href{https://arxiv.org/abs/hep-ph/0202239}{{\ttfamily hep-ph/0202239}}].

\bibitem{Pilaftsis:1997jf}
A.~Pilaftsis, \emph{{CP violation and baryogenesis due to heavy Majorana
  neutrinos}}, \href{https://doi.org/10.1103/PhysRevD.56.5431}{\emph{Phys. Rev.
  D} {\bfseries 56} (1997) 5431}
  [\href{https://arxiv.org/abs/hep-ph/9707235}{{\ttfamily hep-ph/9707235}}].

\bibitem{Pilaftsis:2003gt}
A.~Pilaftsis and T.~E.~J. Underwood, \emph{{Resonant leptogenesis}},
  \href{https://doi.org/10.1016/j.nuclphysb.2004.05.029}{\emph{Nucl. Phys. B}
  {\bfseries 692} (2004) 303}
  [\href{https://arxiv.org/abs/hep-ph/0309342}{{\ttfamily hep-ph/0309342}}].

\bibitem{Pilaftsis:2004xx}
A.~Pilaftsis, \emph{{Resonant tau-leptogenesis with observable lepton number
  violation}}, \href{https://doi.org/10.1103/PhysRevLett.95.081602}{\emph{Phys.
  Rev. Lett.} {\bfseries 95} (2005) 081602}
  [\href{https://arxiv.org/abs/hep-ph/0408103}{{\ttfamily hep-ph/0408103}}].

\bibitem{Pilaftsis:2005rv}
A.~Pilaftsis and T.~E.~J. Underwood, \emph{{Electroweak-scale resonant
  leptogenesis}}, \href{https://doi.org/10.1103/PhysRevD.72.113001}{\emph{Phys.
  Rev. D} {\bfseries 72} (2005) 113001}
  [\href{https://arxiv.org/abs/hep-ph/0506107}{{\ttfamily hep-ph/0506107}}].

\bibitem{Abada:2006fw}
A.~Abada, S.~Davidson, F.-X. Josse-Michaux, M.~Losada and A.~Riotto,
  \emph{{Flavor issues in leptogenesis}},
  \href{https://doi.org/10.1088/1475-7516/2006/04/004}{\emph{JCAP} {\bfseries
  04} (2006) 004} [\href{https://arxiv.org/abs/hep-ph/0601083}{{\ttfamily
  hep-ph/0601083}}].

\bibitem{Abada:2006ea}
A.~Abada, S.~Davidson, A.~Ibarra, F.~X. Josse-Michaux, M.~Losada and A.~Riotto,
  \emph{{Flavour Matters in Leptogenesis}},
  \href{https://doi.org/10.1088/1126-6708/2006/09/010}{\emph{JHEP} {\bfseries
  09} (2006) 010} [\href{https://arxiv.org/abs/hep-ph/0605281}{{\ttfamily
  hep-ph/0605281}}].

\bibitem{Antusch:2006cw}
S.~Antusch, S.~F. King and A.~Riotto, \emph{{Flavour-Dependent Leptogenesis
  with Sequential Dominance}},
  \href{https://doi.org/10.1088/1475-7516/2006/11/011}{\emph{JCAP} {\bfseries
  11} (2006) 011} [\href{https://arxiv.org/abs/hep-ph/0609038}{{\ttfamily
  hep-ph/0609038}}].

\bibitem{Chakraborty:2020gqc}
M.~Chakraborty, R.~Krishnan and A.~Ghosal, \emph{{Predictive $S_4$ flavon model
  with $\text{TM}_1$ mixing and baryogenesis through leptogenesis}},
  \href{https://doi.org/10.1007/JHEP09(2020)025}{\emph{JHEP} {\bfseries 09}
  (2020) 025} [\href{https://arxiv.org/abs/2003.00506}{{\ttfamily
  2003.00506}}].

\bibitem{Fujii:1982ms}
Y.~Fujii, \emph{{Origin of the Gravitational Constant and Particle Masses in
  Scale Invariant Scalar - Tensor Theory}},
  \href{https://doi.org/10.1103/PhysRevD.26.2580}{\emph{Phys. Rev. D}
  {\bfseries 26} (1982) 2580}.

\bibitem{Ford:1987de}
L.~H. Ford, \emph{{COSMOLOGICAL CONSTANT DAMPING BY UNSTABLE SCALAR FIELDS}},
  \href{https://doi.org/10.1103/PhysRevD.35.2339}{\emph{Phys. Rev. D}
  {\bfseries 35} (1987) 2339}.

\bibitem{Chiba:1997ej}
T.~Chiba, N.~Sugiyama and T.~Nakamura, \emph{{Cosmology with x matter}},
  \href{https://doi.org/10.1093/mnras/289.2.L5}{\emph{Mon. Not. Roy. Astron.
  Soc.} {\bfseries 289} (1997) L5}
  [\href{https://arxiv.org/abs/astro-ph/9704199}{{\ttfamily
  astro-ph/9704199}}].

\bibitem{Salati:2002md}
P.~Salati, \emph{{Quintessence and the relic density of neutralinos}},
  \href{https://doi.org/10.1016/j.physletb.2003.07.073}{\emph{Phys. Lett. B}
  {\bfseries 571} (2003) 121}
  [\href{https://arxiv.org/abs/astro-ph/0207396}{{\ttfamily
  astro-ph/0207396}}].

\bibitem{Profumo:2003hq}
S.~Profumo and P.~Ullio, \emph{{SUSY dark matter and quintessence}},
  \href{https://doi.org/10.1088/1475-7516/2003/11/006}{\emph{JCAP} {\bfseries
  11} (2003) 006} [\href{https://arxiv.org/abs/hep-ph/0309220}{{\ttfamily
  hep-ph/0309220}}].

\bibitem{Tsujikawa:2013fta}
S.~Tsujikawa, \emph{{Quintessence: A Review}},
  \href{https://doi.org/10.1088/0264-9381/30/21/214003}{\emph{Class. Quant.
  Grav.} {\bfseries 30} (2013) 214003}
  [\href{https://arxiv.org/abs/1304.1961}{{\ttfamily 1304.1961}}].

\bibitem{Arbey:2008kv}
A.~Arbey and F.~Mahmoudi, \emph{{SUSY constraints from relic density: High
  sensitivity to pre-BBN expansion rate}},
  \href{https://doi.org/10.1016/j.physletb.2008.09.032}{\emph{Phys. Lett. B}
  {\bfseries 669} (2008) 46} [\href{https://arxiv.org/abs/0803.0741}{{\ttfamily
  0803.0741}}].

\bibitem{Arbey:2011gu}
A.~Arbey, A.~Deandrea and A.~Tarhini, \emph{{Anomaly mediated SUSY breaking
  scenarios in the light of cosmology and in the dark (matter)}},
  \href{https://doi.org/10.1007/JHEP05(2011)078}{\emph{JHEP} {\bfseries 05}
  (2011) 078} [\href{https://arxiv.org/abs/1103.3244}{{\ttfamily 1103.3244}}].

\bibitem{DEramo:2017gpl}
F.~D'Eramo, N.~Fernandez and S.~Profumo, \emph{{When the Universe Expands Too
  Fast: Relentless Dark Matter}},
  \href{https://doi.org/10.1088/1475-7516/2017/05/012}{\emph{JCAP} {\bfseries
  05} (2017) 012} [\href{https://arxiv.org/abs/1703.04793}{{\ttfamily
  1703.04793}}].

\bibitem{DiMarco:2018bnw}
A.~Di~Marco, G.~Pradisi and P.~Cabella, \emph{{Inflationary scale, reheating
  scale, and pre-BBN cosmology with scalar fields}},
  \href{https://doi.org/10.1103/PhysRevD.98.123511}{\emph{Phys. Rev. D}
  {\bfseries 98} (2018) 123511}
  [\href{https://arxiv.org/abs/1807.05916}{{\ttfamily 1807.05916}}].

\bibitem{Catena:2009tm}
R.~Catena, N.~Fornengo, M.~Pato, L.~Pieri and A.~Masiero, \emph{{Thermal Relics
  in Modified Cosmologies: Bounds on Evolution Histories of the Early Universe
  and Cosmological Boosts for PAMELA}},
  \href{https://doi.org/10.1103/PhysRevD.81.123522}{\emph{Phys. Rev. D}
  {\bfseries 81} (2010) 123522}
  [\href{https://arxiv.org/abs/0912.4421}{{\ttfamily 0912.4421}}].

\bibitem{Rehagen:2014vna}
T.~Rehagen and G.~B. Gelmini, \emph{{Effects of kination and scalar-tensor
  cosmologies on sterile neutrinos}},
  \href{https://doi.org/10.1088/1475-7516/2014/06/044}{\emph{JCAP} {\bfseries
  06} (2014) 044} [\href{https://arxiv.org/abs/1402.0607}{{\ttfamily
  1402.0607}}].

\bibitem{Meehan:2015cna}
M.~T. Meehan and I.~B. Whittingham, \emph{{Dark matter relic density in
  scalar-tensor gravity revisited}},
  \href{https://doi.org/10.1088/1475-7516/2015/12/011}{\emph{JCAP} {\bfseries
  12} (2015) 011} [\href{https://arxiv.org/abs/1508.05174}{{\ttfamily
  1508.05174}}].

\bibitem{Catena:2004ba}
R.~Catena, N.~Fornengo, A.~Masiero, M.~Pietroni and F.~Rosati, \emph{{Dark
  matter relic abundance and scalar - tensor dark energy}},
  \href{https://doi.org/10.1103/PhysRevD.70.063519}{\emph{Phys. Rev. D}
  {\bfseries 70} (2004) 063519}
  [\href{https://arxiv.org/abs/astro-ph/0403614}{{\ttfamily
  astro-ph/0403614}}].

\bibitem{Damour:1996xx}
T.~Damour, \emph{{Gravitation, experiment and cosmology}},  in \emph{{Les
  Houches Summer School on Gravitation and Quantizations, Session 57}}, 6,
  1996, \href{https://arxiv.org/abs/gr-qc/9606079}{{\ttfamily gr-qc/9606079}}.

\bibitem{Angelantonj:2002ct}
C.~Angelantonj and A.~Sagnotti, \emph{{Open strings}},
  \href{https://doi.org/10.1016/S0370-1573(02)00273-9}{\emph{Phys. Rept.}
  {\bfseries 371} (2002) 1}
  [\href{https://arxiv.org/abs/hep-th/0204089}{{\ttfamily hep-th/0204089}}].

\bibitem{Blanchet:2006dq}
S.~Blanchet and P.~Di~Bari, \emph{{Leptogenesis beyond the limit of
  hierarchical heavy neutrino masses}},
  \href{https://doi.org/10.1088/1475-7516/2006/06/023}{\emph{JCAP} {\bfseries
  06} (2006) 023} [\href{https://arxiv.org/abs/hep-ph/0603107}{{\ttfamily
  hep-ph/0603107}}].

\bibitem{Samanta:2019yeg}
R.~Samanta and M.~Sen, \emph{{Flavoured leptogenesis and CP$^{\mu\tau}$
  symmetry}}, \href{https://doi.org/10.1007/JHEP01(2020)193}{\emph{JHEP}
  {\bfseries 01} (2020) 193}
  [\href{https://arxiv.org/abs/1908.08126}{{\ttfamily 1908.08126}}].

\bibitem{Chen:2019etb}
S.-L. Chen, A.~Dutta~Banik and Z.-K. Liu, \emph{{Leptogenesis in fast expanding
  Universe}}, \href{https://doi.org/10.1088/1475-7516/2020/03/009}{\emph{JCAP}
  {\bfseries 03} (2020) 009}
  [\href{https://arxiv.org/abs/1912.07185}{{\ttfamily 1912.07185}}].

\bibitem{Chang:2021ose}
Z.-F. Chang, Z.-X. Chen, J.-S. Xu and Z.-L. Han, \emph{{FIMP Dark Matter from
  Leptogenesis in Fast Expanding Universe}},
  \href{https://doi.org/10.1088/1475-7516/2021/06/006}{\emph{JCAP} {\bfseries
  06} (2021) 006} [\href{https://arxiv.org/abs/2104.02364}{{\ttfamily
  2104.02364}}].

\bibitem{Konar:2020vuu}
P.~Konar, A.~Mukherjee, A.~K. Saha and S.~Show, \emph{{A dark clue to seesaw
  and leptogenesis in a pseudo-Dirac singlet doublet scenario with
  (non)standard cosmology}},
  \href{https://doi.org/10.1007/JHEP03(2021)044}{\emph{JHEP} {\bfseries 03}
  (2021) 044} [\href{https://arxiv.org/abs/2007.15608}{{\ttfamily
  2007.15608}}].

\bibitem{Mahanta:2019sfo}
D.~Mahanta and D.~Borah, \emph{{TeV Scale Leptogenesis with Dark Matter in
  Non-standard Cosmology}},
  \href{https://doi.org/10.1088/1475-7516/2020/04/032}{\emph{JCAP} {\bfseries
  04} (2020) 032} [\href{https://arxiv.org/abs/1912.09726}{{\ttfamily
  1912.09726}}].

\bibitem{Covi:1996wh}
L.~Covi, E.~Roulet and F.~Vissani, \emph{{CP violating decays in leptogenesis
  scenarios}}, \href{https://doi.org/10.1016/0370-2693(96)00817-9}{\emph{Phys.
  Lett. B} {\bfseries 384} (1996) 169}
  [\href{https://arxiv.org/abs/hep-ph/9605319}{{\ttfamily hep-ph/9605319}}].

\bibitem{Maki:1962mu}
Z.~Maki, M.~Nakagawa and S.~Sakata, \emph{{Remarks on the unified model of
  elementary particles}}, \href{https://doi.org/10.1143/PTP.28.870}{\emph{Prog.
  Theor. Phys.} {\bfseries 28} (1962) 870}.

\bibitem{Kobayashi:1973fv}
M.~Kobayashi and T.~Maskawa, \emph{{CP Violation in the Renormalizable Theory
  of Weak Interaction}}, \href{https://doi.org/10.1143/PTP.49.652}{\emph{Prog.
  Theor. Phys.} {\bfseries 49} (1973) 652}.

\bibitem{Casas:2001sr}
J.~A. Casas and A.~Ibarra, \emph{{Oscillating neutrinos and $\mu \to e,
  \gamma$}}, \href{https://doi.org/10.1016/S0550-3213(01)00475-8}{\emph{Nucl.
  Phys. B} {\bfseries 618} (2001) 171}
  [\href{https://arxiv.org/abs/hep-ph/0103065}{{\ttfamily hep-ph/0103065}}].

\bibitem{Kolb:1990vq}
E.~W. Kolb and M.~S. Turner, \emph{{The Early Universe}}, vol.~69. 1990,
  \href{https://doi.org/10.1201/9780429492860}{10.1201/9780429492860}.

\bibitem{Luty:1992un}
M.~A. Luty, \emph{{Baryogenesis via leptogenesis}},
  \href{https://doi.org/10.1103/PhysRevD.45.455}{\emph{Phys. Rev. D} {\bfseries
  45} (1992) 455}.

\bibitem{Cyburt:2015mya}
R.~H. Cyburt, B.~D. Fields, K.~A. Olive and T.-H. Yeh, \emph{{Big Bang
  Nucleosynthesis: 2015}},
  \href{https://doi.org/10.1103/RevModPhys.88.015004}{\emph{Rev. Mod. Phys.}
  {\bfseries 88} (2016) 015004}
  [\href{https://arxiv.org/abs/1505.01076}{{\ttfamily 1505.01076}}].

\bibitem{arfken}
G.~Arfken, H.~Weber and F.~E. Harris, \emph{{Mathematical Methods for
  Physicists}}. 7th~ed., 2011.

\bibitem{dlmf-nist}
R.~Paris, ``{\it Incomplete Gamma and Related Functions}.''
  \url{https://dlmf.nist.gov/8}.

\bibitem{Strumia:2006db}
A.~Strumia and F.~Vissani, \emph{{Neutrino masses and mixings and...}},
  \href{https://arxiv.org/abs/hep-ph/0606054}{{\ttfamily hep-ph/0606054}}.

\bibitem{tHooft:1976rip}
G.~'t~Hooft, \emph{{Symmetry Breaking Through Bell-Jackiw Anomalies}},
  \href{https://doi.org/10.1103/PhysRevLett.37.8}{\emph{Phys. Rev. Lett.}
  {\bfseries 37} (1976) 8}.

\bibitem{tHooft:1976snw}
G.~'t~Hooft, \emph{{Computation of the Quantum Effects Due to a
  Four-Dimensional Pseudoparticle}},
  \href{https://doi.org/10.1103/PhysRevD.14.3432}{\emph{Phys. Rev. D}
  {\bfseries 14} (1976) 3432}.

\end{thebibliography}\endgroup
\end{document}